**Title:** Spatially-resolved Brillouin spectroscopy reveals biomechanical changes in early ectatic corneal disease and post-crosslinking *in vivo*

**Authors:** Peng Shao[1†], Amira M. Eltony[1†], Theo G. Seiler[1,3,4], Behrouz Tavakol[1], Roberto Pineda[2], Tobias Koller[3], Theo Seiler[3*], Seok-Hyun Yun[1,5**]

[1] Harvard Medical School and Wellman Center for Photomedicine, Massachusetts General Hospital, Boston, MA 02114, USA

[2] Massachusetts Eye and Ear Infirmary, Boston, MA 02114, USA

[3] Institute for Refractive and Ophthalmic Surgery (IROC), Zürich 8002, Switzerland

[4] Universitätsklinik für Augenheilkunde, Inselspital, Bern 3010, Switzerland

[5] Harvard-MIT Health Sciences and Technology, Cambridge, MA 02139, USA

† equal contribution

* theo@seiler.tv, ** syun@hms.harvard.edu

**Abstract:** Mounting evidence connects the biomechanical properties of tissues to the development of eye diseases such as keratoconus, a common disease in which the cornea thins and bulges into a conical shape. However, measuring biomechanical changes *in vivo* with sufficient sensitivity for disease detection has proved challenging. Here, we present a first large-scale study (~200 subjects, including normal and keratoconus patients) using Brillouin light-scattering microscopy to measure longitudinal modulus in corneal tissues with high sensitivity and spatial resolution. Our results *in vivo* provide evidence of biomechanical inhomogeneity at the onset of keratoconus and suggest that biomechanical asymmetry between the left and right eyes may presage disease development. We additionally measure the stiffening effect of corneal crosslinking treatment *in vivo* for the first time. Our results demonstrate the promise of Brillouin microscopy for diagnosis and treatment of keratoconus, and potentially other diseases.

## INTRODUCTION

In parts of the eye, mechanical structure and function are closely linked. The cornea is a prime example; to maintain the prolate configuration that is critical for good vision, the cornea should have adequate mechanical stiffness to counteract external mechanical stresses such as in-plane tension and intraocular pressure (IOP)[1]. Mounting evidence suggests that the biomechanical properties of ocular tissues can be diagnostic targets due to their association with various eye diseases and refractive errors[2]. The mechanical properties of the cornea stem from the intricate lattice of macromolecules such as collagen fibers and proteoglycans making up the corneal stroma[3,4]. Disintegration of the structural organization alters the biomechanical properties and shifts mechanical homeostasis, which, if severe and focal, can result in corneal pathology and impair vision[5,6].

Keratoconus (KC) is a common disease in which the normal prolate cornea locally thins and bulges into a conical shape[7]. Its etiology is not fully understood, but numerous experimental studies have suggested the importance of biomechanical interactions: genetic and molecular studies revealed links to the disintegration of collagen extracellular matrix[8], through increase of matrix metalloproteinase[9] and decrease of lysyl oxidase activity[10]; eye rubbing, a proven risk factor for KC, has been shown to elevate the tear concentration of protease[11]; disrupted collagen orientation[12–14] and reduced mechanical moduli[1] have been observed in KC explants and corneas that developed ectasia after laser-assisted in situ keratomileusis (LASIK).

Current diagnosis of KC is primarily focused on detecting abnormal morphological features in thickness and curvature. Advances in pachymetry and topography have greatly improved the diagnosis and treatment monitoring of KC. However, the current morphological criteria do not allow physicians to predict the rate of progression of KC, a critical input for planning optimal treatment. Early-stage KC prior to the appearance of definitive morphological abnormality is considered a major risk factor for post refractive-surgery ectasia[15,16]. However, detecting such subclinical or forme fruste KC remains an elusive goal. These continuing unmet clinical needs have motivated scientists to look for more sophisticated morphology-based metrics, specific genetic and molecular markers, and other diagnostic approaches such as corneal biomechanics[2,17].

Interestingly, corneal biomechanics has already been embraced in the treatment of KC. Corneal crosslinking (CXL) is the current standard to treat mild and moderate KC. CXL aims to reinforce the three-dimensional macromolecular organization consisting of collagen and proteoglycans and thereby stiffen the cornea[18]. CXL has proven to be clinically effective in most cases, capable of arresting further thinning and protrusion and even improving vision through keratometric flattening of the cones[19–24]. Definitive diagnosis of progressive KC in early stages is critical to achieve optimal therapeutic results[25]. Unfortunately, because current diagnostics cannot differentiate progressive versus non-progressive KC, physicians rely on repeated measurements over several months to monitor the progression of morphological features, often at the risk of having suboptimal therapeutic outcomes due to delayed treatment[26].

With compelling reasons to develop biomechanics-based diagnostics, scientists have devised noninvasive tools capable of measuring biomechanical parameters in patients[17]. Recently, two commercial instruments, the ocular response analyzer (ORA, Reichert Technologies Inc., USA) and the corneal visualization Scheimpflug technology (Corvis ST, OCULUS Optikgeräte GmbH, Germany), became available, which use a single air-puff to induce millimeter-scale deformation of the cornea and extract various viscoelastic parameters[27]. Numerous clinical studies have demonstrated the potential of the technique for cases of moderate to severe KC[28–35]. However, the studies also revealed limitations of the current instruments in identifying mild and moderate KC[28,29]. The rather disappointing results so far are attributed to the fact that corneal deformation is inherently dependent on IOP and the anatomical geometry of the connective tissue shell making up the eye, including both cornea and sclera[36]. More sophisticated computational methods are being sought to deal with the compounding factors and to discern subtle differences in biomechanical properties with improved sensitivity and accuracy[35,37,38]. Another drawback of the air-puff technique is that it measures overall corneal stiffness so would be less sensitive to focal changes in a specific region of the cornea. In research laboratories, other promising techniques are under development, such as optical coherence elastography[39–42] using various external stimuli[43–48]. However, their safety and feasibility *in vivo* has yet to be tested.

We have previously demonstrated Brillouin light-scattering microscopy, which makes it possible to measure longitudinal mechanical modulus in tissues[49]. Unlike air-puff-based methods, the technology does not involve corneal deformation and can interrogate a specific region and depth in corneal tissue with optical resolution[50]. We have established the safety of this technology in humans and reported Brillouin elasticity maps of corneas in advanced KC patients, providing *in vivo* evidence of spatially heterogeneous, biomechanical weakening in late-stage KC corneas[51].

In this paper, we describe the largest clinical Brillouin studies to date, focusing on the early stages of KC. To detect subtle biomechanical changes in mild and moderate KC, we improved the measurement sensitivity and repeatability of Brillouin systems. A careful analysis of Brillouin frequency shifts measured *in vivo* led to several findings about previously unknown, biomechanical features in KC corneas and corneas after CXL treatment, as well as normal corneas. We present results that point to promising diagnostic metrics based on spatially-resolved biomechanical measurements.

## RESULTS

### Development of *in vivo* Brillouin microscopes

We have built two Brillouin microscopes with nearly identical optical designs. One tabletop system was used for studies at MGH in Boston (**Fig. 1A**). The other rack-mounted system was built at MGH and shipped to Zürich for studies at IROC (**Fig. 1B**). Compared to our previous instruments[51], the new systems have several hardware and software upgrades to improve measurement sensitivity, stability and eye tracking, as well as mechanical robustness. Each system consists of a light source, human interface, spectrometer, and computer (**Fig. 1C**). The light source is a single-frequency tunable laser with its output spectrum locked to a near-infrared wavelength of ~780 nm and filtered using an etalon[52]. The laser light is coupled via a polarization-maintaining fiber to the human interface, in which polarization optics route the laser beam to the eye and direct backscattered light to a single-mode fiber. The spectrometer employs two-stage VIPA etalons and apodization filters to achieve a free-spectral range (FSR) of 16 GHz, a resolution of ~120 MHz, and an extinction efficiency of –65 dB. The optical power on the cornea is 3-5 mW (**Fig. 1D**), which is several times lower than the maximum permissible exposure level according to American National Standard Institutes (ANSI) guidelines (*Supplementary Information*). With this optical power applied to corneal tissues, an electron multiplying charge-coupled device (EMCCD) camera recorded a total of ~6,000 Brillouin photons per second. The Brillouin frequency shift, which is half the frequency difference between the Stokes and Anti-Stokes spectral peaks, is determined by curve fitting the recorded Brillouin spectra (**Figs. 1E** and **S2**). With an EMCCD integration time of 0.2 s, we obtained a shot-noise-limited sensitivity of ~ ± 8 MHz in Brillouin frequency shift measurement (**Fig. S4**). The spatial resolution was measured to be ~5 and 30 μm in the transverse and axial directions, respectively (**Fig. S5**). The system employs temperature sensors and temperature-calibrated reference materials to attain short- and long-term repeatability of ~ ± 10 MHz (**Fig. S7**). For corneal scans, an operator adjusts the human interface using a manual joystick to locate the focus at a desired location in front of the cornea while monitoring an eye-tracking video camera based on pupil detection (**Fig. 1F**). At each location, the operator starts a motorized axial scan to obtain a depth profile and calculate the mean Brillouin shift over the stroma (**Fig. 1G**). A typical 2D Brillouin elasticity map constructed from the mean Brillouin shifts measured at various scan locations using standard spatial interpolation is shown in **Fig. 1H**. Please refer to *Materials and Methods* and *Supplementary Information* for more technical details of the systems and data acquisition.

**Brillouin analysis of normal corneas *in vivo***

To establish a baseline for the normal population, we scanned healthy subjects with corneal thickness in a normal range of 495 to 600 μm, no irregular astigmatism, and no history of ophthalmic pathology or surgery. For each subject, a mean central Brillouin shift was obtained from 5 axial scans in a central region within a 2-mm radius from the pupil center. The measured values from 47 healthy subjects (age: 39 ± 13 y/o, 51% female, one eye per subject with left/right chosen at random) ranged from 5.69 to 5.76 GHz with a standard deviation (SD) of 24 MHz. When analyzed as a function of age (**Fig. 2A**), the Brillouin shift values tend to increase with age at a nominal slope of ~3 MHz per decade, but the correlation was not statistically significant. 37 of the 47 subjects had both of their eyes scanned. The mean Brillouin shifts were nearly identical within the instrument's resolution for the left and right eyes: 5.721 ± 0.017 versus 5.723 ± 0.017 GHz (**Fig. 2B**). The difference between the left and right eyes of each individual subject makes a narrow, Gaussian-like distribution centered at 1.8 MHz (mean) (**Fig. 2C**). The bilateral symmetry between the two eyes of healthy subjects is remarkable and in contrast with the large interpersonal variability of ± 24 MHz.

**Brillouin measurements of keratoconus patients at different stages of severity**

To investigate KC, subjects were scanned who have been diagnosed with KC by corneal specialists. The patients were grouped into 4 stages of severity according to the Amsler-Krumeich classification[53] using quantitative criteria, namely refractive error, best spectacle-corrected visual acuity, minimum corneal thickness, and maximum curvature (K-max). For each cornea, pachymetry and topography images identified a "cone" region, which we define as the area within a 1 mm radius from the thinnest point. Brillouin scans were conducted at 20 to 40 lateral locations to generate a Brillouin map of the cornea. A few representative Brillouin maps are shown in **Fig. 3**, comparison to Brillouin maps of normal corneas (**Fig. 3A**). For advanced KC corneas in stage III or IV, Brillouin maps showed considerable non-uniformity compared to normal corneas (**Fig. 3C**). The cones, defined herein by the thinnest points, are approximately coincident with the bulging areas of abnormal curvature and surface elevation, and these areas are characterized by low Brillouin frequency shift values, distinctly different by as much as 100-200 MHz compared to healthy corneas. Stage-I corneas are characterized by mild morphological anomalies, and the thinnest points do not always overlap with the locations of maximum curvature or posterior elevation. Compared to stage-III-IV cases, state-I corneas exhibit much more uniform Brillouin maps, and the regions of lower Brillouin values are less pronounced (**Fig. 3B**).

The locations of minimum thickness, i.e. cone centers, are distributed predominantly in the center-temporal region (**Fig. 4A**). The XY coordinate in mm of the center of the distribution was (0.45, –0.85) for stages I and II (n=34, 36 ± 11 y/o, 47% female) and (0.48, –1.2) for stages III and IV (n=12, 46 ± 14 y/o, 75% female). The data suggest that the cone location within an individual may not change much as KC progresses. From the Brillouin maps, we calculated the mean Brillouin frequency shifts in the cone regions and plotted the data as a function of minimum corneal thickness (**Fig. 4B**) and maximum anterior sagittal curvature, K max (**Fig. 4C**). Weak to moderate trends of decreasing Brillouin frequency shifts with increasing KC severity were measured, with coefficients of determination for linear regression being $R^2 = 0.40$ for corneal thickness and $R^2 = 0.28$ for K-max. The graph in **Fig. 4D** shows the distributions of Brillouin frequency shift values in cone regions of KC patients in comparison to Brillouin values measured in central regions of normal patients (within a 1-mm radius from the pupil center). While the reduction of Brillouin values is evident for advanced KC, no statistically significant difference was observed between stage-I-II KC and normal corneas. The mean Brillouin values of the stage-I and -II groups are lower than the healthy group by 3 and 7 MHz, respectively. Nonetheless, the observed null correlation results from large interpersonal variability within each group. Notably, the subject-to-subject variation in stage I and II patients is greater than the intrinsic interpersonal difference among healthy subjects.

## Analysis of regional biomechanical variation in keratoconus corneas

The observation of large interpersonal variability in both normal and KC corneas prompted us to seek metrics that are less subject to individual variations. We hypothesized that most of the variability stems from natural personal differences in the composition, collagen organization, and/or hydration level of corneal tissues. We further reasoned that these factors would likely affect Brillouin values uniformly across the entire cornea and therefore can be canceled out in differential metrics, such as regional difference within the cornea. From Brillouin maps, we found that, whereas normal corneas exhibit relatively uniform Brillouin values across their about-4-mm-wide scanned areas (**Fig. 5A**), the Brillouin shifts in KC corneas have a clear tendency to increase linearly with the distance from the cone in all stages, including stages I and II (**Figs. 5B** and **5C**). The slope of regional variation increased with the severity of KC (**Fig. S10**). We compared the mean Brillouin frequency shift in the cone region versus outside-cone region—defined by all scanned points outside a 3-mm radius from the thinnest point—for each cornea. For healthy subjects, central and peripheral Brillouin values are not different (**Fig. 5D**, n=16, 39 ± 10 y/o, 25% female). By contrast, highly significant differences ($p < 0.001$) were found between cone and outside-cone regions for all stages of KC (**Figs. 5E** and **5F**). **Fig. 5G** shows regional variations or the differences of outside- and inside-cone values for all groups. All KC groups exhibit higher regional differences than the normal group. However, the statistical significance is moderate ($p < 0.01$) except for stage IV, with considerable overlaps of Brillouin values with the normal group.

## Analysis of bilateral symmetry in KC corneas

KC is known to be bilateral, but often asymmetric. The high degree of left-to-right symmetry in central Brillouin values in healthy subjects motivated us to analyze bilateral symmetry for KC corneas. Representative Brillouin maps from a patient with stage-I KC in both corneas are shown in **Fig. 6A**. The thickness and posterior elevation maps share symmetric, but not identical, features. Asymmetry in the Brillouin maps is apparent although subtle. To quantify the asymmetry, we calculated the difference in Brillouin frequency shifts between left (OS) and right (OD) eyes in two corresponding locations: near the pupil center and in the cone. The asymmetry in Brillouin values in the central regions of KC corneas was moderately higher ($p < 0.01$, n=4, 38 ± 5 y/o, 50% female) than in normal corneas (**Fig. 6B**). Remarkably, the Brillouin asymmetry between the left and right cone regions, which are identified using the thickness maps, is significantly larger than for the normal group (**Fig. 6B**). The mean bilateral difference in the stage-I cones was 24 MHz, well separated from the normal population that is predominantly below 12 MHz. Despite the relatively small sample size, the data are compelling and reveal that bilateral symmetry may be a promising metric for diagnosis of early-stage KC. The narrow distribution of the data suggests that the compounding factors that are responsible for the subject-to-subject variability in Brillouin values are equally present in both eyes and, therefore, cancelled out in this differential metric.

### Brillouin measurements of KC patients *in vivo* treated with collagen crosslinking

CXL has previously been shown to increase Brillouin shifts in *ex vivo* porcine corneas, but it has remained unknown whether Brillouin shifts are altered following CXL *in vivo* in humans. To address this question, we conducted a cross-sectional study with KC patients who have been selected for and treated with CXL using a standard protocol at IROC in the past 1 to 4 years (n=16, 35 ± 11 y/o, 19% female). As a comparison group, we also recruited CXL-candidates who were eligible for CXL at IROC under consistent selection criteria —progressive, early-stage KC—and scheduled to undergo customized CXL[54] at IROC (n=8, 26 ± 11 y/o, 50% female). The pre-CXL group had significantly lower ($p < 0.001$) Brillouin values (mean: 5.697 ± 0.029 GHz) compared to the central Brillouin shifts in the normal control group (5.722 ± 0.017 GHz**, Fig. 7**). A mild statistical difference ($p < 0.05$) was found between the pre-CXL (5.697 ± 0.029 GHz) and post-CXL (5.725 ± 0.030 GHz) groups in terms of Brillouin frequency shifts in cone regions (**Fig. 7**). Notably, no statistical difference was observed between the post-CXL and normal groups. Taken together, the results are consistent with the prevailing hypothesis that CXL *in vivo* does enhance the mechanical stiffness of weakened KC corneas. They further suggest that the degree of crosslinking induced by the current protocol is such that the corneal stiffness is brought to the normal level of healthy corneas. The SD in the post-CXL group was ± 30 MHz, slightly greater than ± 25 MHz and ± 17 MHz for the pre-CXL and normal groups, respectively. This broader distribution may be due to possible variability in the response to CXL procedures.

## DISCUSSION

The magnitude of the optical frequency shift that we measure is proportional to the velocity of longitudinal acoustic waves in the tissue. The acoustic velocity is in turn related to the square root of longitudinal modulus of elasticity[55]. For viscoelastic materials such as tissues, longitudinal modulus is nearly equal to bulk modulus, or the material's resistance to volume change. Because biological tissues are ~70% water by volume, they are nearly incompressible and exhibit high longitudinal modulus, in the range of 2.5 to 3.5 GPa for most tissues[56], producing Brillouin frequencies of 5.69 to 5.77 GHz in corneas[51] and up to 6.1 GHz in crystalline lenses[57] (at an optical wavelength of 780 nm). The bulk modulus of aqueous humor, which is nearly the same as that of water, is 2.2 GPa, for which we measured a Brillouin shift of 5.25 GHz (Fig. 1G). The higher longitudinal-modulus of corneal tissue compared to water derives from the contribution of the other constituent matter such as ions and proteins, that make the tissue less compressible. Brillouin spectroscopy has been shown to be sensitive to small differences in molecular composition, concentration (or water content)[56], and photochemical crosslinking[58,59]. Brillouin measurement may be less sensitive to processes such as pure polymeric rearrangement or changes that are predominantly entropic, but relatively more sensitive to processes involving changes in molecular concentration and composition.

In contrast to bulk modulus, shear or Young's modulus characterizes a material's response to stress under free boundary conditions. Since soft matter is deformed relatively easily at constant volume, its shear modulus is orders-of-magnitude smaller than its bulk modulus. The shear modulus measured by mechanical methods at low deformation frequencies (< 100 Hz) is typically ~250 kPa for normal corneal tissue. Although longitudinal and shear moduli are in principle two independent quantities and can have vastly different magnitudes, empirical data obtained with polymers, hydrogels, and various tissues including corneas have shown reasonably high, quantitative correlation between Brillouin frequency shift and quasi-static shear modulus[50,51,60]. This measurement result suggests that in practice for corneal tissues, natural or pathologic processes that cause shear modulus to increase (or decrease) are likely to cause Brillouin shifts to increase (or decrease).

**Figure 8** illustrates a physical model of the cornea and how KC and CXL may influence different moduli of elasticity. The mechanical properties of corneal stroma are determined predominantly by the density and arrangement of collagen fibrils, which account for about 70% of the total corneal dry weight and are interconnected by proteoglycan and glycosaminoglycan (GAG) chains[3,4,61]. We expect that longitudinal modulus will decrease as the tissue degenerates in KC and that CXL can increase longitudinal modulus through accompanying molecular compaction (not directly by entropic contribution of crosslinks). Considering the typical relationship of density and refractive index (*Methods*), Brillouin frequency shift should change in the same direction as longitudinal modulus.

The Brillouin data from normal adult corneas *in vivo* showed a modest age-dependence with a slope of ~3 MHz/decade. Unexpectedly, we observed relatively large subject-to-subject differences (± 24 MHz SD) even among age-matched subjects (Fig. 2A). Given the interpersonal variability, the rather small age dependence is surprising, as corneal surgeons often experience less soft and pliable corneas in old patients compared to young. Numerous *ex vivo* mechanical measurements have reported increasing trends of Shear and Young's moduli with age[62–64]. About 12% increase of Young's modulus per decade has been estimated[64]. Employing an often-used empirical log-log conversion factor of 60:1, the slope of 12% per decade in Young's modulus corresponds to, approximately, 0.2% per decade in longitudinal modulus, or 12 MHz/decade in Brillouin frequency (*Supplementary Information*). Previously, electron microscopy has revealed thickening of collagen fibrils[65] and decrease of their inter-fibrillar spacing[66] with age. Such changes in microstructure[67], as well as age-related increase of glycation-mediated collagen crosslinking[68], are thought to elevate corneal stiffness over age[67]. The apparent discrepancy between our measurement and previous *ex vivo* measurements (as well as surgeons' perception by *in vivo* palpation) has yet to be resolved. It is interesting to note that *in vivo* measurements using ORA found no increase or even small decline of corneal hysteresis and corneal resistance factor with age[33,69–71].

As an aside, ultrasound pachymetry—a current gold standard for corneal thickness measurement—assumes constant acoustic speed for all subjects and then calculates thickness from the time delay of ultrasound pulses reflected from two corneal surfaces[72]. The accuracy of commercial ultrasound pachymeters is known to be < ± 1%. Based on our Brillouin data, we expect hypersonic-wave-based thickness measurement to have an error of ~ ± 0.4% (± 24 MHz / 5.72 GHz) associated with the interpersonal and age-dependent variability.

We supposed that the interpersonal variability stems from natural differences in individual's normal physiological conditions. For example, Brillouin shift is sensitive to the water content in hydrogels and, therefore, hydration level of corneal tissues[73]. The hydration level is known to differ individually and fluctuate throughout the day[74]. This hydration variation is likely to contribute to the measured interpersonal variability. Brillouin shift of tissue can also be sensitive to body temperature, as expected from a temperature dependence of 7.4 MHz/°C for water (Fig. S6). In the future, it would be possible to record body temperature and apply a correction factor to measured Brillouin values. Since these physiological factors affect globally the entire cornea, we had expected much smaller interpersonal variability for differential metrics. Indeed, we measured substantially smaller "intra-personal" differences between left and right eyes (Fig. 2B) and between the central and peripheral regions of each cornea (Fig. 5D).

Variability in corneal biomechanics across individuals can introduce significant errors in applanation-tonometry IOP readings[75]. Hence, individual differences in Brillouin value may be used to apply a "biomechanics correction" to improve the accuracy of IOP measurements. Individual differences in Brillouin value may also provide useful input for refractive treatments: it is known that corneal hydration affects the excimer laser ablation rate in LASIK surgeries[76,77], and personal biomechanical differences are thought to explain the variability in refractive outcomes following cataract and astigmatic correction keratotomy (AK) surgeries[78]. It may be possible to individually tailor ablation parameters or AK nomograms based on Brillouin measurements made prior to treatment.

In the present study of KC patients, we confirmed our previous finding that the cones in advanced KC corneas show up clearly in Brillouin maps as the regions with significantly reduced Brillouin shifts (Fig. 3B)[51]. However, the correspondence between Brillouin maps and pachymetry and topography is less clear in early-stage KC corneas, and Brillouin shifts correlate weakly with morphological parameters such as thickness and curvature (Figs. 4B and C). We found no statistical difference between stage-I-II KC patients and healthy subjects (Fig. 4D). The Brillouin values within stage-I and II groups each exhibited larger inter-subject variation than the normal group. The large variations might originate from the physiological factors discussed above and, possibly, additional factors associated with KC pathogenesis.

The distinction between early-stage KC and normal groups was more pronounced in the analysis of regional differences between cone and outside-cone regions. In most stage-I and -II subjects and all stage-III and IV subjects, the cone regions had lower Brillouin shifts than the peripheral regions. This supports the long-held hypothesis that the degeneration of corneal tissues occurs focally at the cones[17]. Given moderate statistical significance ($p < 0.01$) between stage-I and normal values, the regional-difference parameter has potential to be a useful metric to characterize early-stage KC. However, the two-point comparison alone is unlikely to be diagnostic of KC, considering the overlap of data between different groups (Fig. 5G). With more accurate, comprehensive mapping of the cornea in future, more sophisticated metrics based on specific spatial patterns in Brillouin maps may be developed.

The most promising metric we have found is the difference in Brillouin shifts between left and right eyes. The bilateral asymmetry measured in stage-I KC patients is clearly separated from that of normal corneas, and reflects the clinical experience that in many keratoconus patients one eye shows a more advanced stage of disease relative to the other[79]. It warrants more investigation of this metric in a larger scale study, which may include patients with only one eye in stage-I KC and a subclinical or KC suspect contralateral eye. The bilateral asymmetry indices of healthy subjects form a normal distribution with width limited by the system's sensitivity. With enhanced Brillouin signals through longer data acquisition time, the distribution in the healthy population may turn out to be narrower. The bilateral asymmetry index may prove useful for identifying patients at an earlier stage or those at risk of developing KC. If so, this index could be used to develop new diagnostic criteria in combination with, or even beyond the current morphology-based gold standard. It is worthwhile to note that it is the bilateral asymmetry measured at the thinnest point that showed the sharp distinction. The asymmetry index measured at the pupil center showed only moderate statistical difference from the normal population (Fig. 6B). This consistently supports the hypothesis of focal weakening in the development of KC[17].

Besides corneas, other ocular tissues such as the crystalline lens and sclera are thought to present a high degree of symmetry in their healthy states. In fact, in our previous study of crystalline lenses *in vivo*, we have measured large interpersonal variability, as much as ± 60 MHz in age-matched groups, as opposed to small differences between left and right lenses within the measurement uncertainty of ± 10 MHz[57]. Brillouin-based bilateral asymmetry has potential to be a general, useful technique for the detection of subtle pathological anomalies and drug-induced changes in the biomechanical properties of the lens and sclera.

We have also presented the first *in vivo* Brillouin data from patients after clinical CXL treatment. The mean Brillouin shift of the post-CXL group was 30 MHz higher than the pre-CXL group, providing evidence for the positive effect of CXL on the stiffening of corneal stroma. However, the statistical significance of this finding was modest ($p < 0.05$), in part due to the large interpersonal variation in, particularly, the post-CXL group. Previously we have observed a depth-dependent, dramatic increase of Brillouin shifts, as much as a few hundred MHz, in porcine and bovine corneas immediately after *ex vivo* CXL[58]. Judging from the *in vivo* data, the dramatic increases measured *ex vivo* were likely exaggerated, presumably by CXL-induced tissue compaction and hydration changes—possible artifacts in *ex vivo* experiments. Our data clearly showed milder post-CXL increase *in vivo*, presumably because of the regulation of hydration level and remodeling of collagen fibrillar structure. Numerous studies have been conducted using ORA and Corvis ST instruments to measure the biomechanical effects of CXL both *ex vivo* and *in vivo*, so far producing informative but mostly inconsistent results. For example, several *in vivo* studies found no significant change in the magnitude of corneal deformation with air puffs following CXL[80–82], whereas in *ex vivo* samples CXL clearly resulted in reduced corneal deformation[83,84], suggestive of stiffening. Our data represents the first *in vivo* indication of post-CXL corneal stiffening.

The observation of large interpersonal variation following CXL treatment motivates future longitudinal studies in which KC patients are scanned pre- and post-CXL and the differential changes in Brillouin shifts are monitored over time. The time-lapse data will have much lower, self-canceled interpersonal variation and may provide quantitative, mechanistic insights into the effects and variability of CXL procedures.

The current Brillouin systems require a data acquisition time of > 0.2 s per depth point and ~12 s for a single axial scan. In clinical practice, this speed would be acceptable for acquiring data at only a few locations, such as the thinnest point and a peripheral point, which would be sufficient for measuring the regional difference and bilateral asymmetry. However, improvement of speed will permit more comprehensive mapping of the entire cornea (~9 mm diameter) in a short time acceptable in clinical settings. The total number of Brillouin scattered photons that were collected and analyzed at the spectrometer was about $3 \times 10^{-13}$ of the number of laser photons incident on the corneal surface. The theoretical maximum collection efficiency of Brillouin light scattering in an on-axis confocal configuration is estimated to be $\frac{2\pi^2}{\lambda^3}\frac{(n^2-1)^2}{K}k_BT$, where $k_BT = 4.23 \times 10^{-21}$ J at the cornea temperature, $n$ is refractive index, $K$ is bulk modulus, and $\lambda$ is optical wavelength (*Supplementary Information*). For corneal tissues, the efficiency is ~ $5 \times 10^{-11}$. There is room to increase the Brillouin signal by as much as 10-fold by reducing optical loss in the detection path.

In conclusion, these clinical results have demonstrated the diagnostic potential of Brillouin light scattering spectroscopy for corneal examinations, particularly for detection and treatment monitoring of KC patients. The two most promising metrics for KC-discrimination are based on bilateral asymmetry and focal weakening in early-stage KC patients, a finding made possible by harnessing the intrinsic ability of Brillouin technology to measure local biomechanical properties of tissues with high spatial resolution and sensitivity. Our results suggest that spatially-resolving Brillouin spectroscopy may prove useful not only for the diagnosis of KC but also for improving CXL treatment[54], and potentially in many other applications, including screening subjects at risk of ectasia prior to refractive keratectomy surgeries[15,16,85] and optimizing corneal incision parameters in cataract and astigmatic correction keratotomy surgeries by taking individual biomechanics into account[78]. Brillouin measurements may also be used to improve the accuracy of basic examinations such as intraocular pressure measurement[75].

## MATERIALS AND METHODS

### Study design

The objective of this study was to assess the feasibility of using Brillouin spectroscopy for clinical characterization of human corneal biomechanics. Specifically, to evaluate the potential of Brillouin spectroscopy for diagnosis and treatment of keratoconus. For this purpose, we examined the interpersonal variability, age-dependence, regional heterogeneity, and bilateral asymmetry in Brillouin values among normal corneas, keratoconus corneas of different severity, and corneas with history of collagen crosslinking treatment.

The general inclusion criteria were subjects with clear enough cornea or clear enough media to permit imaging, and who fulfill all specific inclusion/exclusion criteria of each study group. Excluded from this study were monocular volunteers. The normal cornea group included healthy subjects with normal appearing corneas respecting all the general inclusion/exclusion criteria, with less than ± 3 diopters refractive error, normal corneal thickness (495 to 600 μm), normal corneal topography (no asymmetric or irregular astigmatism, no skewed axis), no corneal pathology, and no history of eye diseases, except presbyopia and/or cataract. The keratoconus group included subjects with irregular corneas determined by distorted keratometry mires or irregularities in Scheimpflug photography, and slit-lamp biomicroscopic signs such as Vogt's striae or Fleischer's ring or corneal scarring consistent with keratoconus. The CXL-treated group included patients having undergone CXL surgery within the past 4 years and excluded corneas with scarring or having undergone additional ocular surgeries, such as laser vision correction surgeries.

Sample sizes for each group were chosen in advance based on a power analysis. We used preliminary data to estimate the sample size required to establish a correlation between corneal stiffness and progression of the disease, subject age, or treatment with CXL, with > 90% power.

Studies were performed at two locations: at the Institute for Refractive and Ophthalmic Surgery (IROC) in Zürich, Switzerland, and at Massachusetts Eye and Ear Infirmary (MEEI) and Massachusetts General Hospital (MGH) in Boston, USA, following approval from the Institutional Review Board (IRB) of Partners HealthCare, the Partners Human Research Committee (PHRC) and the Institutional Review Board of IROC, Zürich. Informed consent was obtained from every patient before imaging. The clinical trial "A Study to Test the Potential of Brillouin Microscopy for Biomechanical Properties Measurements in Human Cornea" is registered at www.clinicaltrials.gov (National Clinical Trial Identifier: NCT03220529).

**Brillouin data acquisition**

During Brillouin measurement, the subject sits with their chin and forehead resting in the human interface headrest, directing their gaze towards a fixation target. For healthy subjects with normal corneas, 5 axial scans are taken within a ∅2 mm zone in the central cornea. To create a Brillouin map of the cornea, 20 – 40 axial scans are taken at different locations laterally across the cornea, usually within the central ∅6 mm zone. Light power was calibrated to be 5 mW on the cornea surface. Each axial scan comprises 40 points separated by a step size of 30 μm, spanning air to cornea to aqueous humor. The EMCCD exposure time for each step is typically 300 ms, resulting in a total axial scan time of about 12 s. Following each axial scan, temperature-corrected calibration spectra are taken using known reference materials (polystyrene and water) to determine the dispersion rate (in GHz/pixel) of the spectrum pattern acquired from the cornea, and thus the corresponding Brillouin frequency shift (Fig. S3). From the calibrated axial scan data, points acquired within the corneal stroma are extracted and averaged to yield a single value for each lateral scan location (Fig. 1G). Linear interpolation between these lateral measurement points is then used to create the final 2D Brillouin map (Fig. 1H). Data processing was conducted using MATLAB (Mathworks, Inc.).

**Corneal topographies**

The corneal topographies were generated using a commercial topographer (Vario, Wavelight, Erlangen, Germany) and a commercial Scheimpflug camera (Pentacam HR 70700, OCULUS, Wetzlar, Germany). Patients were asked not to use contact lenses in the two weeks prior to each examination.

**Corneal collagen crosslinking**

Patients receiving CXL suffered from a documented progressive, primary keratoconus. Excluded from this study were eyes with previous ocular surgery, penetrating trauma, glaucoma, aphakia, endothelial cell count less than 2300 cells/mm$^2$, corneas with no distinctive posterior float area, pellucid marginal degeneration, corneal scars that may interfere with Scheimpflug photography, history of recurrent erosions, neurodermatitis and connective tissue diseases, or pregnancy and breast feeding. Eyes treated before October 2016 received standard CXL (Dresden protocol, 5.4 J/cm$^2$ using 9 mW/cm$^2$) including a 9.0 mm circular epithelium debridement and application of 0.1% riboflavin in 16% dextran T500 for 30 minutes. After October 2016, customized CXL of the cornea was performed using an irradiance of 15 mW/cm² and a variable radiant exposure (Mosaic, Avedro Inc., Waltham, MA)[54].

**Data analysis**

To process a single axial scan, we first determine the rate of spectral dispersion using the included calibration spectrum (see Fig. S3). With this information, we can then determine the Brillouin frequency shift at each of the (typically) 40 scan points spanning air to cornea to aqueous humor. We then segment this Brillouin shift versus axial depth data, based on Brillouin value and other characteristic signal features, in order to identify and extract only points within the corneal stroma. Finally, we average the Brillouin shift measured across all stroma points to obtain a single Brillouin value for the axial scan.

To create a Brillouin map of the cornea, we use standard linear interpolation to connect Brillouin values from 20-40 axial scans at different lateral locations. To analyze the spatial variation in Brillouin values across the cornea, we define various regions of interest (see Fig. S9 for a graphical illustration of these zones). The central region is defined as the zone within a 1-mm radius around the pupil center; the peripheral region is the area with a distance > 3 mm from the pupil center. In keratoconic corneas, the 'cone region' is defined as the area within a 1 mm radius of the thinnest point (the 'cone center') identified using corneal topography, and the 'outside-cone region' refers to the area > 3 mm from the cone center.

**Statistical analysis**

Custom MATLAB software was used for data processing. Statistical analysis was carried out using the Statistics toolbox in MATLAB. To examine the significance of differences observed in the mean Brillouin values of different groups of subjects (such as comparing normal versus KC stage I) we used unpaired, two-tailed *t*-tests. To evaluate the significance of differences observed within a single cornea (such as comparing one region to another) we used paired, two-tailed *t*-tests. Data were considered significant if p-values were less than 0.05 (95% confidence intervals).

## ACKNOWLEDGEMENTS

We acknowledge Dr. Giuliano Scarcelli and Dr. Sebastien Besner for their contributions during the development of the Brillouin systems, Dr. Valéry Wittwer for assistance with the age-dependence study, and Dr. Dominik Beck for helping coordinate the IROC trial. The project was funded by the U.S. National Institutes of Health (R01EY025454, P41EB015903) and the Harvard Catalyst Incubator program (UL1-RR025758).

## SUPPLEMENTARY MATERIALS

Fig. S1: Brillouin light scattering in the cornea.

Fig. S2: Quantitative relationship between Brillouin modulus (frequency shift) and Shear modulus.

Fig. S3: Spectrometer frequency calibration.

Fig. S4: Signal to noise ratio and measurement sensitivity.

Fig. S5. Axial resolution.

Fig. S6: Temperature dependence of Brillouin measurements and calibration.

Fig. S7. Measurement of short-term and long-term stability.

Fig. S8. Daily variation of *in vivo* measurements.

Fig. S9. Definition of zones for analyzing spatial variation.

Fig. S10. Regional differences in Brillouin frequency shift measured in corneas.

Table. S1. Technical specifications of the Brillouin Systems

## REFERENCES AND NOTES

1. Dupps, W. J. & Wilson, S. E. Biomechanics and wound healing in the cornea. *Exp. Eye Res.* **83,** 709–720 (2006).
2. Girard, M. J. A. *et al.* Translating Ocular Biomechanics into Clinical Practice: Current State and Future Prospects. *Curr. Eye Res.* **40,** 1–18 (2015).
3. Maurice, D. in *The Eye: Vegetative Physiology and Biochemistry* (ed. Davson, H.) 1–58 (Academic Press, Orlando, FL, 1984).
4. Kamma-Lorger, C. S. *et al.* Collagen and mature elastic fibre organisation as a function of depth in the human cornea and limbus. *J. Struct. Biol.* **169,** 424–430 (2010).
5. McMonnies, C. W. & Boneham, G. C. Corneal responses to intraocular pressure elevations in keratoconus. *Cornea* **29,** 764–70 (2010).
6. Roy, A. S. & Dupps, W. J. Patient-specific computational modeling of keratoconus progression and differential responses to collagen cross-linking. *Investig. Ophthalmol. Vis. Sci.* **52,** 9174–9187 (2011).
7. Rabinowitz, Y. S. Keratoconus. *Surv. Ophthalmol.* **42,** 297–319 (1998).
8. Lu, Y. *et al.* Genome-wide association analyses identify multiple loci associated with central corneal thickness and keratoconus. *Nat. Genet.* **45,** 155–163 (2013).
9. Seppala, H. P. S. *et al.* EMMPRIN and MMP-1 in keratoconus. *Cornea* **25,** 325–330 (2006).
10. Dudakova, L. *et al.* Changes in lysyl oxidase (LOX) distribution and its decreased activity in keratoconus corneas. *Exp. Eye Res.* **104,** 74–81 (2012).
11. Balasubramanian, S. A., Pye, D. C. & Willcox, M. D. P. Effects of eye rubbing on the levels of protease, protease activity and cytokines in tears: Relevance in keratoconus. *Clin. Exp. Optom.* **96,** 214–218 (2013).
12. Meek, K. M. *et al.* Changes in collagen orientation and distribution in keratoconus corneas. *Investig. Ophthalmol. Vis. Sci.* **46,** 1948–1956 (2005).
13. Morishige, N. *et al.* Second-harmonic imaging microscopy of normal human and keratoconus cornea. *Investig. Ophthalmol. Vis. Sci.* **48,** 1087–1094 (2007).
14. Dawson, D. G. *et al.* Corneal ectasia after excimer laser keratorefractive surgery:

histopathology, ultrastructure, and pathophysiology. *Ophthalmology* **115,** 2181–2191 (2008).
15. Seiler, T. & Quurke, A. W. Iatrogenic keratectasia after LASIK in a case of forme fruste keratoconus. *J. Cataract Refract. Surg.* **24,** 1007–1009 (1998).
16. Rabinowitz, Y. Ectasia after laser in situ keratomileusis. *Curr. Opin. Ophthalmol.* **17,** 421–427 (2006).
17. Roberts, C. J. & Dupps, W. J. Biomechanics of corneal ectasia and biomechanical treatments. *J. Cataract Refract. Surg.* **40,** 991–998 (2014).
18. Spoerl, E., Huhle, M. & Seiler, T. Induction of Cross-links in Corneal Tissue. *Exp. Eye Res.* **66,** 97–103 (1998).
19. Caporossi, A., Baiocchi, S., Mazzotta, C., Traversi, C. & Caporossi, T. Parasurgical therapy for keratoconus by riboflavin-ultraviolet type A rays induced cross-linking of corneal collagen: preliminary refractive results in an Italian study. *J. Cataract Refract. Surg.* **32,** 837–45 (2006).
20. Raiskup-Wolf, F., Hoyer, A., Spoerl, E. & Pillunat, L. E. Collagen crosslinking with riboflavin and ultraviolet-A light in keratoconus: Long-term results. *J. Cataract Refract. Surg.* **34,** 796–801 (2008).
21. Caporossi, A. *et al.* Long-term Results of Riboflavin Ultraviolet A Corneal Collagen Cross-linking for Keratoconus in Italy: The Siena Eye Cross Study. *Am. J. Ophthalmol.* **149,** 585–593 (2010).
22. Pron, G., Ieraci, L. & Kaulback, K. *Collagen Cross-Linking Using Riboflavin and Ultraviolet-A for Corneal Thinning Disorders: An Evidence-Based Analysis*. *Ontario Health Technology Assessment Series* **11,** (2011).
23. Hersh, P. S., Greenstein, S. A. & Fry, K. L. Corneal collagen crosslinking for keratoconus and corneal ectasia: One-year results. *J. Cataract Refract. Surg.* **37,** 149–160 (2011).
24. Wittig-Silva, C. *et al.* A Randomized, Controlled Trial of Corneal Collagen Cross-Linking in Progressive Keratoconus. *Ophthalmology* **121,** 812–821 (2014).
25. Wollensak, G. Crosslinking treatment of progressive keratoconus: new hope. *Curr. Opin. Ophthalmol.* **17,** 356–60 (2006).
26. Li, X., Yang, H. & Rabinowitz, Y. S. Longitudinal study of keratoconus progression. *Exp. Eye Res.* **85,** 502–507 (2007).
27. Luce, D. A. Determining in vivo biomechanical properties of the cornea with an ocular response analyzer. *J. Cataract Refract. Surg.* **31,** 156–162 (2005).
28. Kirwan, C., O'Malley, D. & O'Keefe, M. Corneal hysteresis and corneal resistance factor in keratoectasia: Findings using the Reichert Ocular Response Analyzer. *Ophthalmologica* **222,** 334–337 (2008).
29. Fontes, B. M., Ambrósio, R. J., Jardim, D., Velarde, G. C. & Nosé, W. Corneal Biomechanical Metrics and Anterior Segment Parameters in Mild Keratoconus. *Ophthalmology* **117,** 673–679 (2010).
30. Fontes, B. M., Ambrósio, R., Velarde, G. C. & Nosé, W. Corneal biomechanical evaluation in healthy thin corneas compared with matched keratoconus cases. *Arq. Bras. Oftalmol.* **74,** 13–16 (2011).
31. Touboul, D. *et al.* Early biomechanical keratoconus pattern measured with an ocular response analyzer: Curve analysis. *J. Cataract Refract. Surg.* **37,** 2144–50 (2011).
32. Alió, J. L. *et al.* Keratoconus-integrated characterization considering anterior corneal aberrations, internal astigmatism, and corneal biomechanics. *J. Cataract Refract. Surg.* **37,**

552–568 (2011).
33. Ortiz, D., Piñero, D., Shabayek, M. H., Arnalich-Montiel, F. & Alió, J. L. Corneal biomechanical properties in normal, post-laser in situ keratomileusis, and keratoconic eyes. *J. Cataract Refract. Surg.* **33,** 1371–5 (2007).
34. Shah, S., Laiquzzaman, M., Bhojwani, R., Mantry, S. & Cunliffe, I. Assessment of the Biomechanical Properties of the Cornea with the Ocular Response Analyzer in Normal and Keratoconic Eyes. *Investig. Ophthalmol. Vis. Sci.* **48,** 3026–3031 (2007).
35. Galletti, J. G., Pförtner, T. & Bonthoux, F. F. Improved Keratoconus Detection by Ocular Response Analyzer Testing After Consideration of Corneal Thickness as a Confounding Factor. *J. Refract. Surg.* **28,** 202–8 (2012).
36. Kling, S. & Marcos, S. Contributing factors to corneal deformation in air puff measurements. *Investig. Ophthalmol. Vis. Sci.* **54,** 5078–5085 (2013).
37. Roberts, C. J. *et al.* Factors Influencing Corneal Deformation and Estimation of Intraocular Pressure. *Investig. Ophthalmol. Vis. Sci.* **52,** 4384 (2011).
38. Metzler, K. M. *et al.* Deformation response of paired donor corneas to an air puff: intact whole globe versus mounted corneoscleral rim. *J. Cataract Refract. Surg.* **40,** 888–96 (2014).
39. Ford, M. R., Dupps, W. J., Rollins, A. M., Roy, A. S. & Hu, Z. Method for optical coherence elastography of the cornea. *J. Biomed. Opt.* **16,** 16005 (2011).
40. Dorronsoro, C., Pascual, D., Pérez-Merino, P., Kling, S. & Marcos, S. Dynamic OCT measurement of corneal deformation by an air puff in normal and cross-linked corneas. *Biomed. Opt. Express* **3,** 473 (2012).
41. Wang, S. & Larin, K. V. Optical coherence elastography for tissue characterization: A review. *J. Biophotonics* **8,** 279–302 (2015).
42. Singh, M. *et al.* Investigating Elastic Anisotropy of the Porcine Cornea as a Function of Intraocular Pressure With Optical Coherence Elastography. *J. Refract. Surg.* **32,** 562–567 (2016).
43. Wang, H. An ultrasonic technique for the measurement of the elastic moduli of human cornea. *J. Biomech.* **29,** 1633–1636 (1996).
44. Tanter, M., Touboul, D., Gennisson, J. L., Bercoff, J. & Fink, M. High-resolution quantitative imaging of cornea elasticity using supersonic shear imaging. *IEEE Trans. Med. Imaging* **28,** 1881–1893 (2009).
45. He, X. & Liu, J. A quantitative ultrasonic spectroscopy method for noninvasive determination of corneal biomechanical properties. *Investig. Ophthalmol. Vis. Sci.* **50,** 5148–5154 (2009).
46. Wang, S. & Larin, K. V. Shear wave imaging optical coherence tomography (SWI-OCT) for ocular tissue biomechanics. *Opt. Lett.* **39,** 41 (2014).
47. Touboul, D. *et al.* Supersonic shear wave elastography for the in vivo evaluation of transepithelial corneal collagen cross- linking. *Investig. Ophthalmol. Vis. Sci.* **55,** 1976–1984 (2014).
48. Nguyen, T.-M. *et al.* Visualizing ultrasonically induced shear wave propagation using phase-sensitive optical coherence tomography for dynamic elastography. *Opt. Lett.* **39,** 838–841 (2014).
49. Scarcelli, G. & Yun, S. H. Confocal Brillouin microscopy for three-dimensional mechanical imaging. *Nat. Photonics* **2,** 39–43 (2008).
50. Scarcelli, G., Pineda, R. & Yun, S. H. Brillouin optical microscopy for corneal

biomechanics. *Investig. Ophthalmol. Vis. Sci.* **53,** 185–190 (2012).
51. Scarcelli, G., Besner, S., Pineda, R., Kalout, P. & Yun, S. H. In Vivo Biomechanical Mapping of Normal and Keratoconus Corneas. *JAMA Ophthalmol.* **133,** 480–482 (2015).
52. Shao, P., Besner, S., Zhang, J., Scarcelli, G. & Yun, S.-H. Etalon filters for Brillouin microscopy of highly scattering tissues. *Opt. Express* **24,** 22232–22238 (2016).
53. Amsler, M. Kératocône classique et kératocône fruste; arguments unitaires. *Ophthalmologica* **111,** 96–101 (1946).
54. Seiler, T. G. *et al.* Customized Corneal Cross-linking: One-Year Results. *Am. J. Ophthalmol.* **166,** 14–21 (2016).
55. Randall, J. & Vaughan, J. M. The Measurement and Interpretation of Brillouin Scattering in the Lens of the Eye. *Proc. R. Soc. B* **214,** 449–470 (1982).
56. Scarcelli, G. *et al.* Noncontact three-dimensional mapping of intracellular hydromechanical properties by Brillouin microscopy. *Nat. Methods* **12,** 1132–1134 (2015).
57. Besner, S., Scarcelli, G., Pineda, R. & Yun, S. H. In vivo brillouin analysis of the aging crystalline lens. *Investig. Ophthalmol. Vis. Sci.* **57,** 5093–5100 (2016).
58. Scarcelli, G. *et al.* Brillouin microscopy of collagen crosslinking: Noncontact depth-dependent analysis of corneal elastic modulus. *Investig. Ophthalmol. Vis. Sci.* **54,** 1418–1425 (2013).
59. Kwok, S. J. J. *et al.* Selective two-photon collagen crosslinking in situ measured by Brillouin microscopy. *Optica* **3,** 469 (2016).
60. Scarcelli, G., Kim, P. & Yun, S. H. In vivo measurement of age-related stiffening in the crystalline lens by Brillouin optical microscopy. *Biophys. J.* **101,** 1539–1545 (2011).
61. Komai, Y. & Ushiki, T. The three-dimensional organisation of collagen fibrils in the human cornea and sclera. *Investig. Ophthalmol. Vis. Sci.* **32,** 2244–2258 (1991).
62. Elsheikh, A. *et al.* Assessment of Corneal Biomechanical Properties and Their Variation with Age. *Curr. Eye Res.* **32,** 11–19 (2007).
63. Elsheikh, A., Geraghty, B., Rama, P., Campanelli, M. & Meek, K. M. Characterization of age-related variation in corneal biomechanical properties. *J. R. Soc. Interface* **7,** 1475–1485 (2010).
64. Cartwright, N. E. K., Tyrer, J. R. & Marshall, J. Age-Related Differences in the Elasticity of the Human Cornea. *Investig. Ophthalmol. Vis. Sci.* **52,** 4324–4329 (2011).
65. Daxer, A., Misof, K., Grabner, B., Ettl, A. & Fratzl, P. Collagen Fibrils in the Human Corneal Stroma: Structure and Aging. *Investig. Ophthalmol. Vis. Sci.* **39,** 644–648 (1998).
66. Malik, N. S. *et al.* Ageing of the human corneal stroma: structural and biochemical changes. *Biochim. Biophys. Acta* **1138,** 222–228 (1992).
67. Goh, K. L. *et al.* Ageing Changes in the Tensile Properties of Tendons: Influence of Collagen Fibril Volume Fraction. *J. Biomech. Eng.* **130,** 21011 (2008).
68. Bailey, A. J., Paul, R. G. & Knott, L. Mechanisms of maturation and ageing of collagen. *Mech. Ageing Dev.* **106,** 1–56 (1998).
69. Kotecha, A., Elsheikh, A., Roberts, C. R., Zhu, H. & Garway-Heath, D. F. Corneal Thickness- and Age-Related Biomechanical Properties of the Cornea Measured with the Ocular Response Analyzer. *Investig. Ophthalmol. Vis. Sci.* **47,** 5337–5347 (2006).
70. Kamiya, K., Shimizu, K. & Ohmoto, F. Effect of Aging on Corneal Biomechanical Parameters Using the Ocular Response Analyzer. *J. Refract. Surg.* **25,** 888–893 (2009).
71. Sharifipour, F., Panahi-bazaz, M., Bidar, R., Idani, A. & Cheraghian, B. Age-related

variations in corneal biomechanical properties. *J. Curr. Ophthalmol.* **28,** 117–122 (2016).
72. Swartz, T., Marten, L. & Wang, M. Measuring the cornea: the latest developments in corneal topography. *Curr. Opin. Ophthalmol.* **18,** 325–333 (2007).
73. Silverman, R. H. *et al.* Effect of corneal hydration on ultrasound velocity and backscatter. *Ultrasound Med. Biol.* **35,** 839–846 (2009).
74. Harper, C. L. *et al.* Diurnal variations in human corneal thickness. *Br. J. Ophthalmol.* **80,** 1068–1072 (1996).
75. Liu, J. & Roberts, C. J. Influence of corneal biomechanical properties on intraocular pressure measurement: Quantitative analysis. *J. Cataract Refract. Surg.* **31,** 146–155 (2005).
76. Dougherty, P. J., Wellish, K. L. & Maloney, R. K. Excimer laser ablation rate and corneal hydration. *Am. J. Ophthalmol.* **118,** 169–176 (1994).
77. Kim, W. S. & Jo, J. M. Corneal hydration affects ablation during laser in situ keratomileusis surgery. *Cornea* **20,** 394–7 (2001).
78. Denoyer, A., Ricaud, X., Van Went, C., Labbé, A. & Baudouin, C. Influence of corneal biomechanical properties on surgically induced astigmatism in cataract surgery. *J. Cataract Refract. Surg.* **39,** 1204–1210 (2013).
79. Naderan, M., Rajabi, M. T. & Zarrinbakhsh, P. Intereye asymmetry in bilateral keratoconus, keratoconus suspect and normal eyes and its relationship with disease severity. *Br. J. Ophthalmol.* **101,** 1475–1482 (2017).
80. Mikielewicz, M., Kotliar, K., Barraquer, R. I. & Michael, R. Air-pulse corneal applanation signal curve parameters for the characterisation of keratoconus. *Br. J. Ophthalmol.* **95,** 793–8 (2011).
81. Goldich, Y. *et al.* Clinical and corneal biomechanical changes after collagen cross-linking with riboflavin and UV irradiation in patients with progressive keratoconus: results after 2 years of follow-up. *Cornea* **31,** 609–614 (2012).
82. Bak-Nielsen, S., Pedersen, I. B., Ivarsen, A. & Hjortdal, J. Dynamic Scheimpflug-based Assessment of Keratoconus and the Effects of Corneal Cross-linking. *J. Refract. Surg.* **30,** 408–414 (2014).
83. Kling, S., Bekesi, N., Dorronsoro, C., Pascual, D. & Marcos, S. Corneal viscoelastic properties from finite-element analysis of in vivo air-puff deformation. *PLoS One* **9,** (2014).
84. Bekesi, N., Kochevar, I. E. & Marcos, S. Corneal Biomechanical Response Following Collagen Cross-Linking With Rose Bengal–Green Light and Riboflavin-UV. *Investig. Ophthalmol. Vis. Sci.* **57,** 992–1001 (2016).
85. Binder, P. S. *et al.* Keratoconus and corneal ectasia after LASIK. *J. Cataract Refract. Surg.* **31,** 2035–8 (2005).
86. Cheng, X. & Pinsky, P. M. Mechanisms of self-organization for the collagen fibril lattice in the human cornea. *J. R. Soc. Interface* **10,** 20130512 (2013).
87. Akhtar, S. *et al.* Ultrastructural analysis of collagen fibrils and proteoglycans in keratoconus. *Acta Ophthalmol.* **86,** 764–772 (2008).
88. Akhtar, S., Almubrad, T., Paladini, I. & Mencucci, R. Keratoconus corneal architecture after riboflavin/ultraviolet A cross-linking: ultrastructural studies. *Mol. Vis.* **19,** 1526–37 (2013).

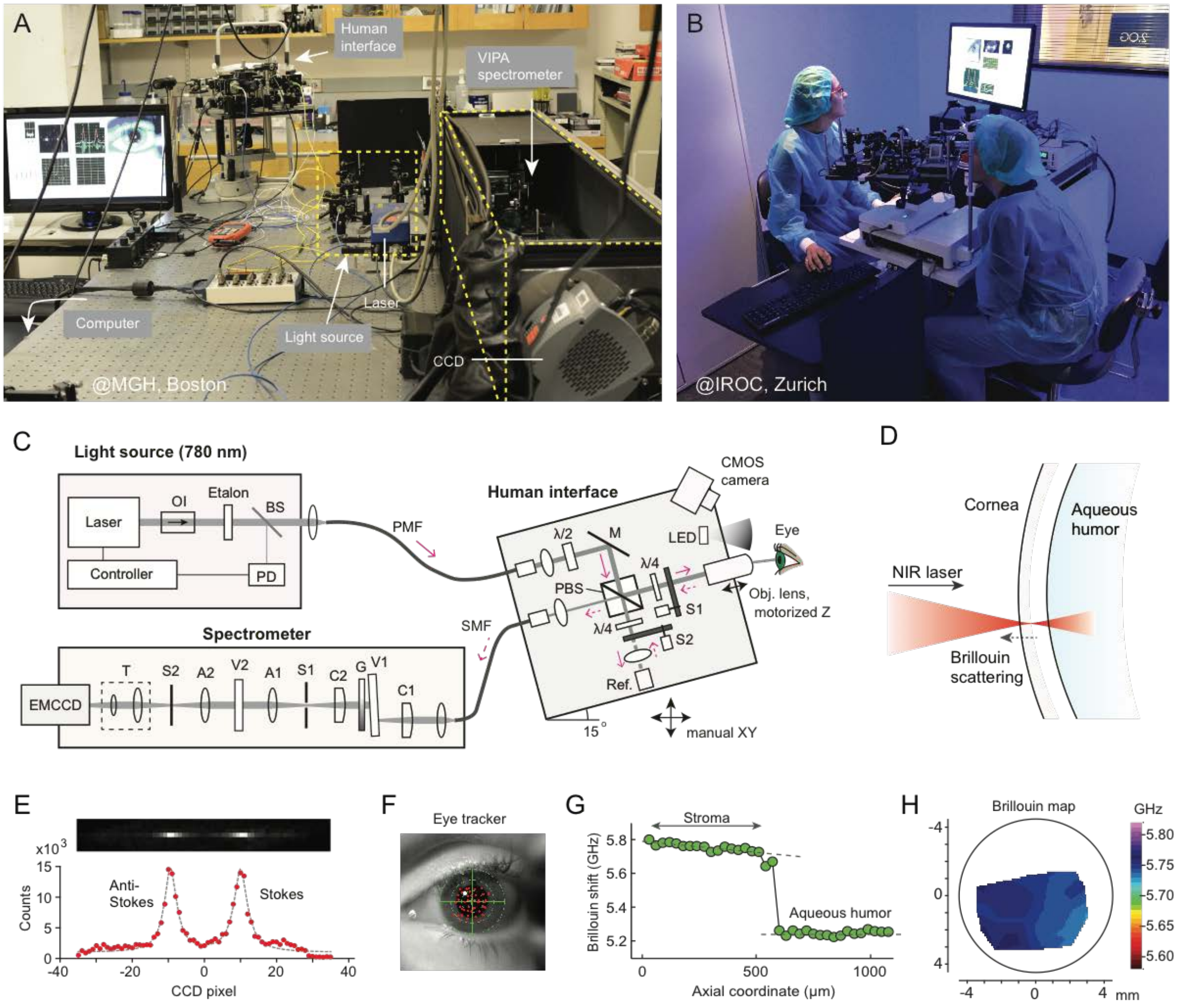

**Fig. 1**. The Brillouin ocular analyzer system. (A) The tabletop Brillouin imaging system in the laboratory at Massachusetts General Hospital in Boston. (B) Brillouin measurement of a volunteer by an operator, using the portable Brillouin ocular scanner in the clinical setting at IROC in Zürich, Switzerland. (C) Schematic of the Brillouin imaging system, composed of three parts: a light source, a human scanning interface built on a modified slit-lamp platform, and a high-resolution, two-stage VIPA spectrometer consisting of two crossed-axis VIPA etalons. OI: optical isolator; PD: photodiode; BS: beam sampler; PMF: polarization maintaining single-mode fiber; SMF: single mode fiber; $\lambda/2$: half waveplate; $\lambda/4$: quarter-waveplate; M: mirror; Obj. L: objective lens; PBS: polarizing beam-splitter; S1/S2: optical shutters; Ref: reference materials; C1/C2: cylindrical lens; A1/A2: achromatic lens; S1/S2: shutters; V1/V2: VIPAs. (D) Probe beam geometry. (E) Representative Brillouin signals from the cornea recorded *in vivo* using the EMCCD camera in the spectrometer. (F) An eye-tracking camera view showing locations of axial scans (red dots). (G) A representative axial (depth) scan profile of measured frequency

shifts across the cornea and anterior chamber of a healthy volunteer. (H) A typical Brillouin elasticity map of a normal cornea.

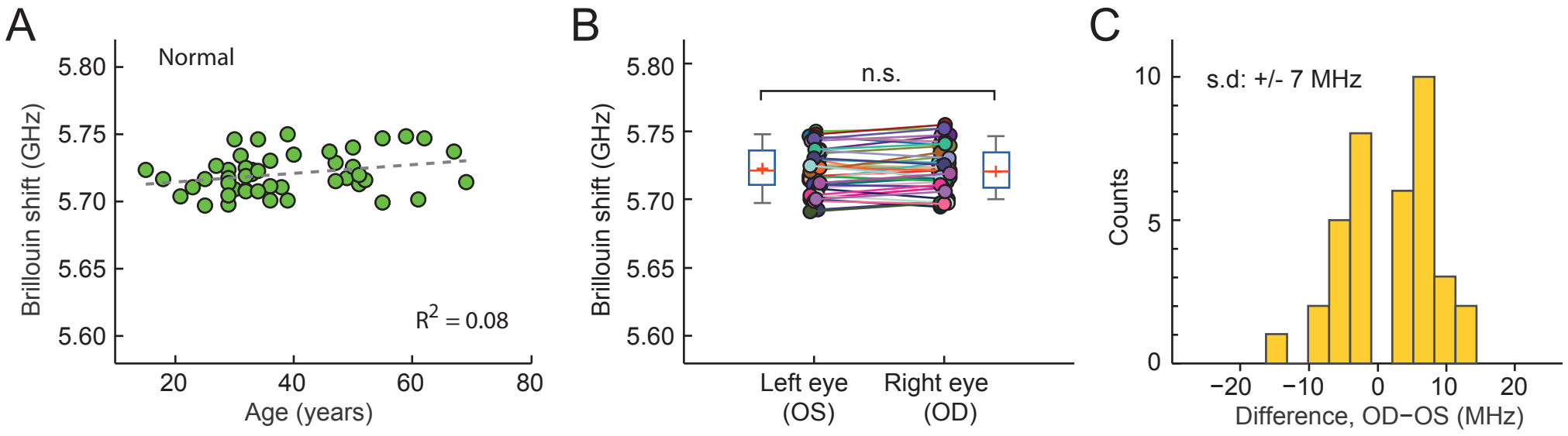

**Fig. 2**. *In vivo* Brillouin measurements of normal corneas. (A) Brillouin frequency shifts measured at corneal centers in healthy subjects of various ages (n=47, 39 ± 13 y/o, 51% female). Broken line depicts a linear regression fit with a slope of 0.3 × age MHz/year, with Pearson's $p$ = 0.06, $r^2$ = 0.08. (B) Comparison of Brillouin shifts between the left (OS) and right (OD) eyes of each subject (n=37). No statistically significant difference is found. (C) Distribution of OD-OS difference (mean: 0.0018 ± 0.007 GHz), from 37 subjects whose eyes were both scanned (38 ± 14 y/o, 60% female).

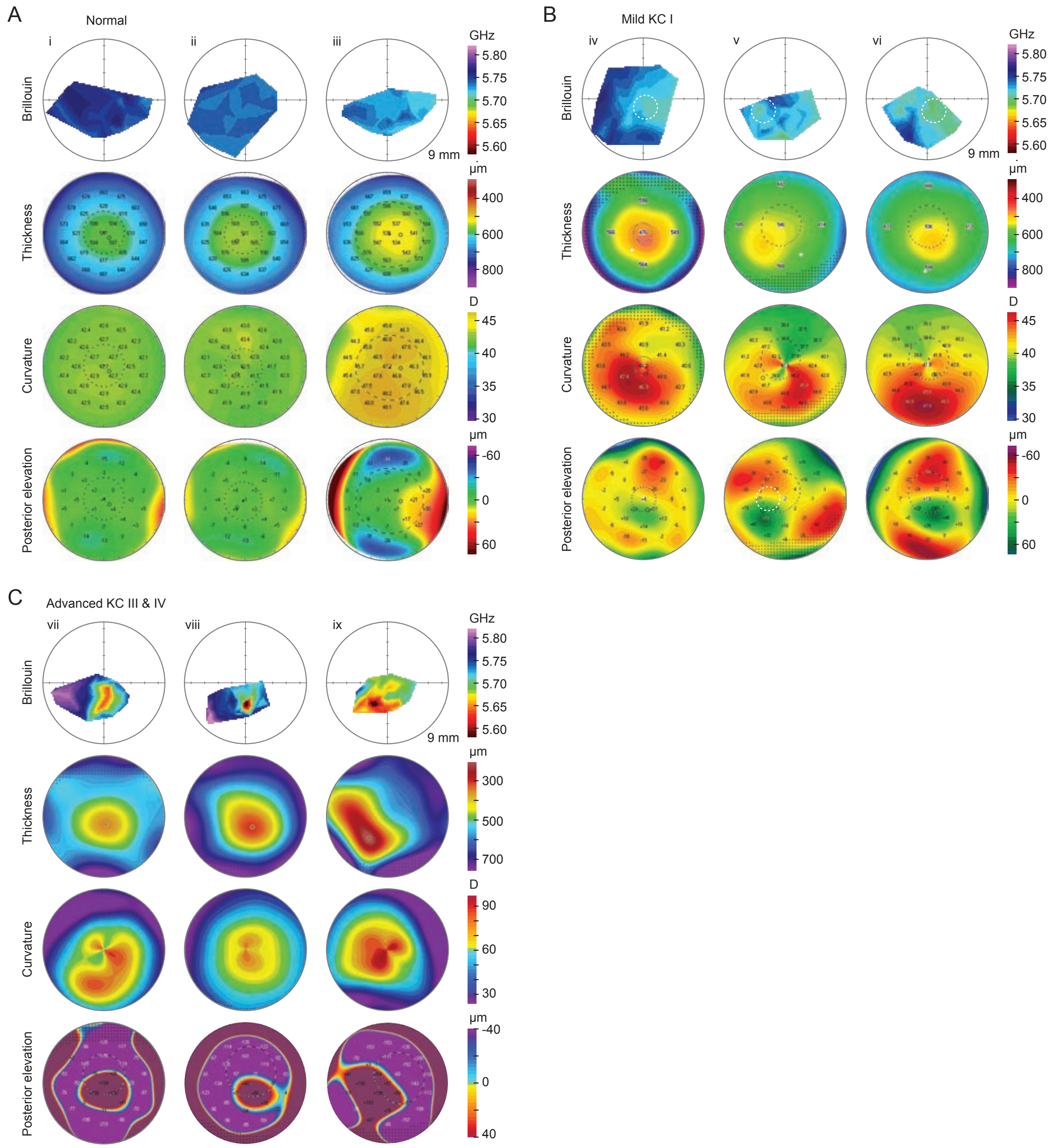

**Fig. 3**. Representative Brillouin images of corneas *in vivo*. (A) normal corneas (subjects i to iii), (B) corneas diagnosed with keratoconus in Stage I (iv-vi), and (C) corneas with severe keratoconus in Stage III or IV (vii-ix). The circles (black) indicate an area of 9 mm in diameter centered at the pupil center. Rows 2-4 are topographical images of the corneas, showing corneal thickness, sagittal curvature, and posterior surface elevation. (Note the difference in color maps and ranges due to difference in clinical Scheimpflug imaging systems.)

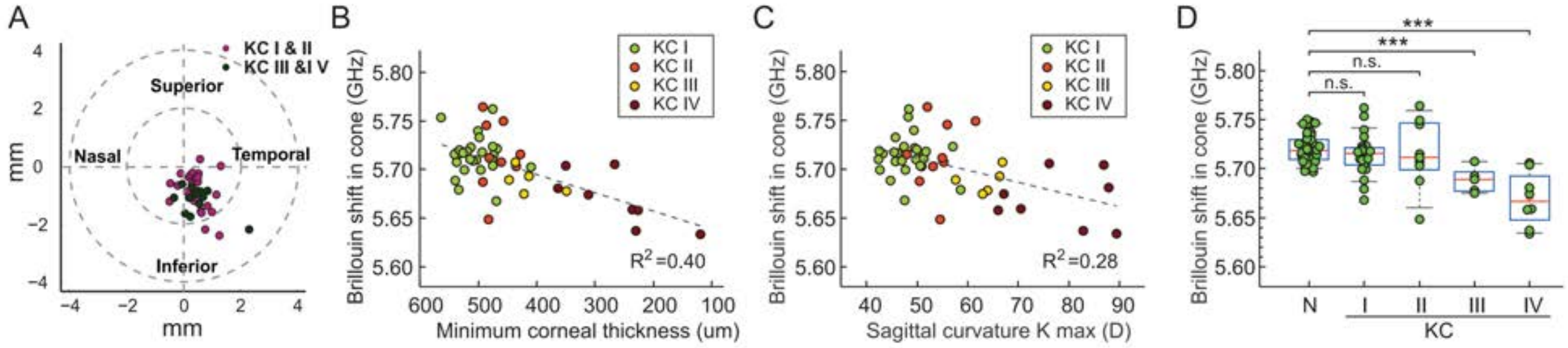

**Fig. 4.** Correlation between Brillouin frequency values and KC severity. (A) Locations of the thinnest points identified by pachymetry in patients diagnosed with KC stage I or II (n=34, 36 ± 11 y/o; 47% female) or KC stage III or IV (n=12, 46 ± 14 y/o; 75% female). (B) Correlation of Brillouin shift with corneal thickness. (C) Correlation of Brillouin shift with sagittal curvature. Dashed lines: linear fits. (D) Comparison of Brillouin frequency values of normal subjects and KC patients at different stages: I (n = 25, 36 ± 11 y/o, 46% female), II (n = 9, 45 ± 13 y/o, 44% female), III (n = 4, 47 ± 12 y/o, 40% female) and IV (n = 8, 45 ± 15 y/o, 25% female). n.s.: non-significant, ***: $p < 0.001$, by unpaired *t*-test.

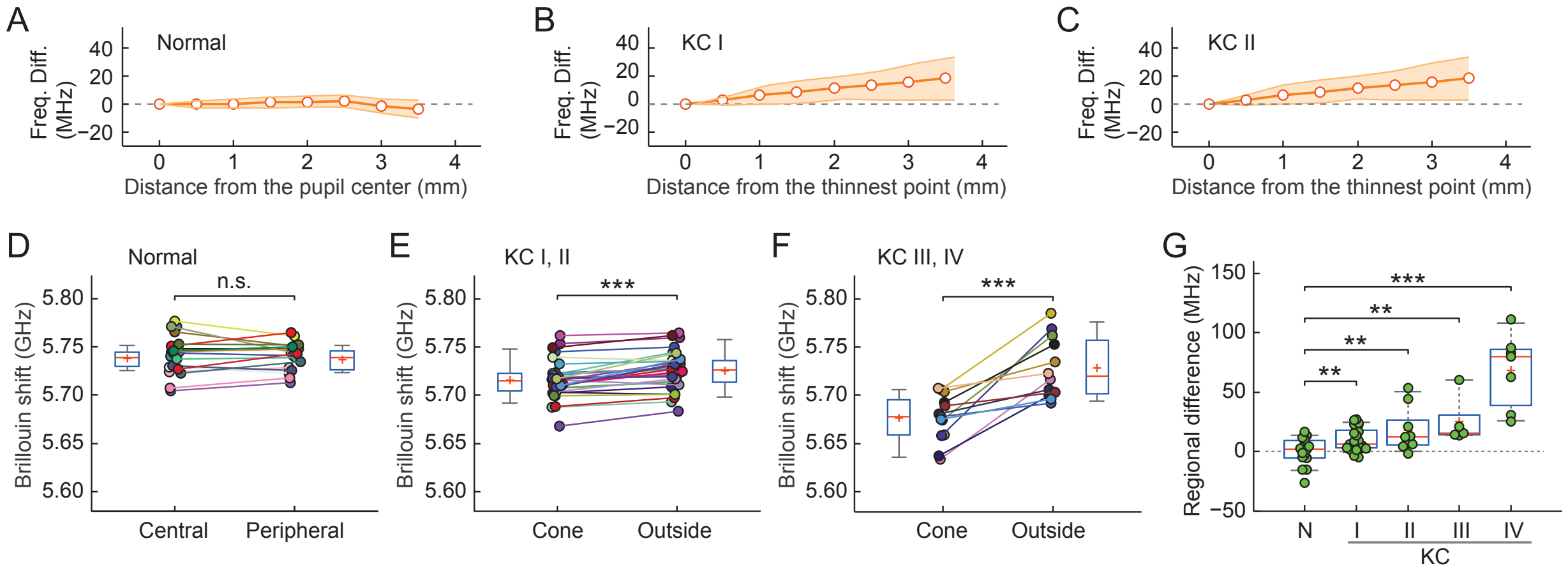

**Fig. 5.** Spatial variation of Brillouin frequency value across corneas. (A-C) Lateral variation of Brillouin shifts in: (A) normal subjects, (B) stage-I KC patients, (C) stage-II KC patients. Circles and shaded regions represent average and SD of 4 to 16 maps. (D-F) Pairwise comparison of Brillouin values in corneal centers or cones to Brillouin values in peripheral regions, for (D) normal subjects (n=16, 39 ± 10 y/o, 25% female), (E) Stage I-II KC, (F) Stage III-IV KC patients. **: $p < 0.01$, ***: $p < 0.001$, by paired $t$-test. (G) Difference in Brillouin shift measured in the outside-cone (peripheral) versus cone (central) regions for KC and normal (N) groups. n.s.: non-significant, **: $p < 0.01$, ***: $p < 0.001$, by unpaired $t$-test.

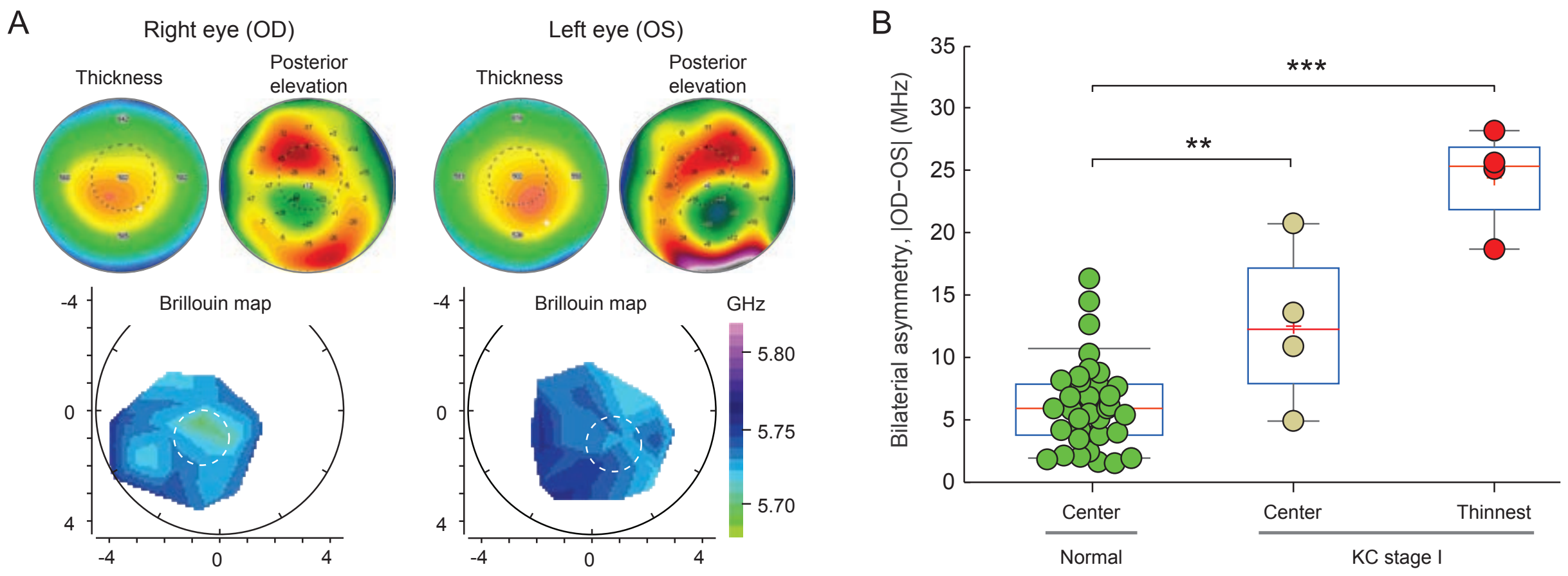

**Fig. 6.** Bilateral asymmetry in early-stage KC. (A) Brillouin, corneal thickness and posterior elevation maps of the right and left eyes of a stage-I KC patient. Broken red circles denote 1-mm-radius cone regions centered at the thinnest points. (B) Absolute difference in Brillouin frequency values between left and right eyes, measured at the corneal centers for normal (green) subjects and stage-I KC patients (beige) and at the thinnest points for the KC patients (n=4, 38 ± 5 y/o, 50% female). **: $p < 0.01$, ***: $p < 0.001$, by unpaired t-test.

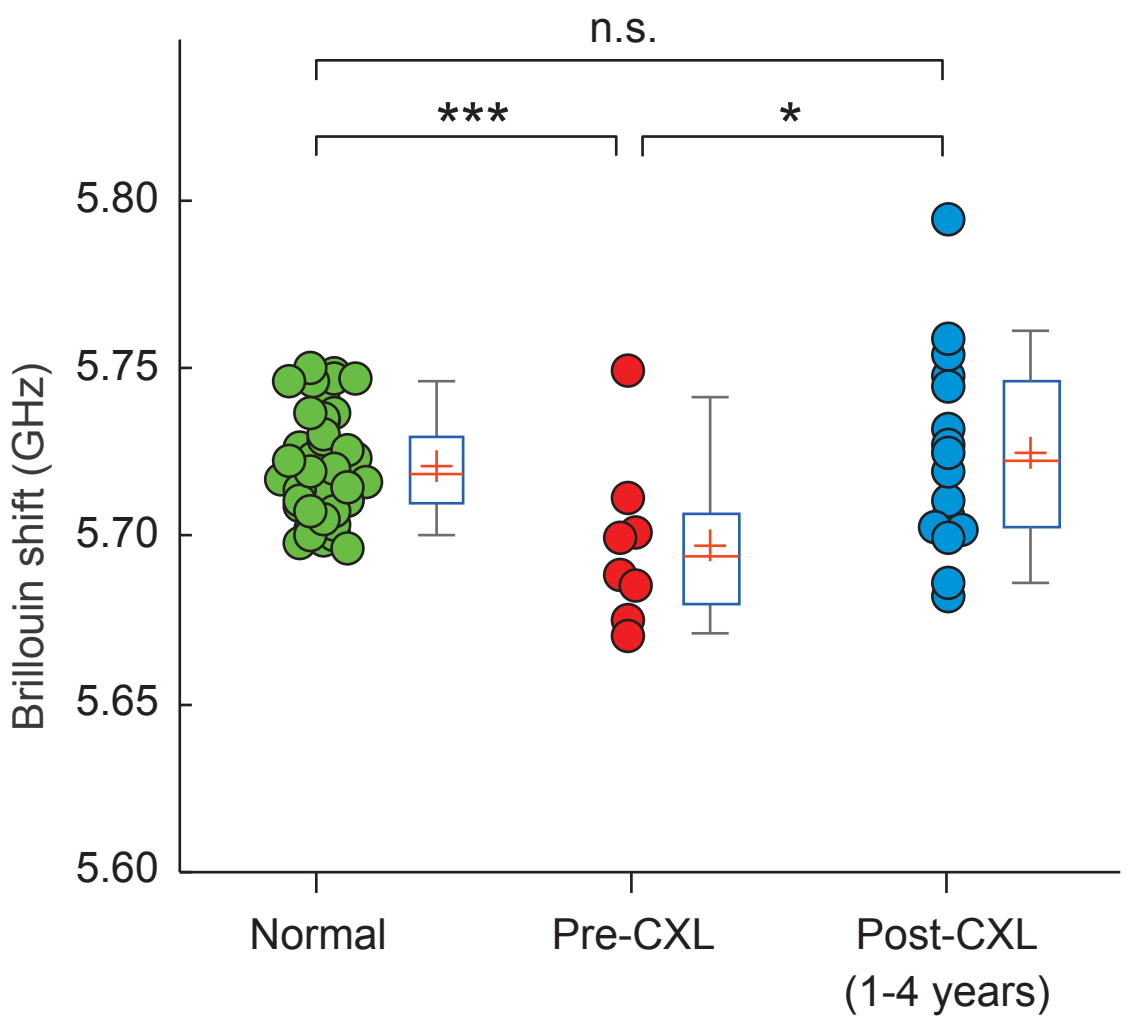

**Fig. 7.** Brillouin analysis of CXL. Brillouin frequencies measured in three groups: normal, CXL candidates prior to CXL, and post-CXL patients within 1 to 4 years after treatment. *: $p < 0.05$, ***: $p < 0.001$, by unpaired *t*-test.

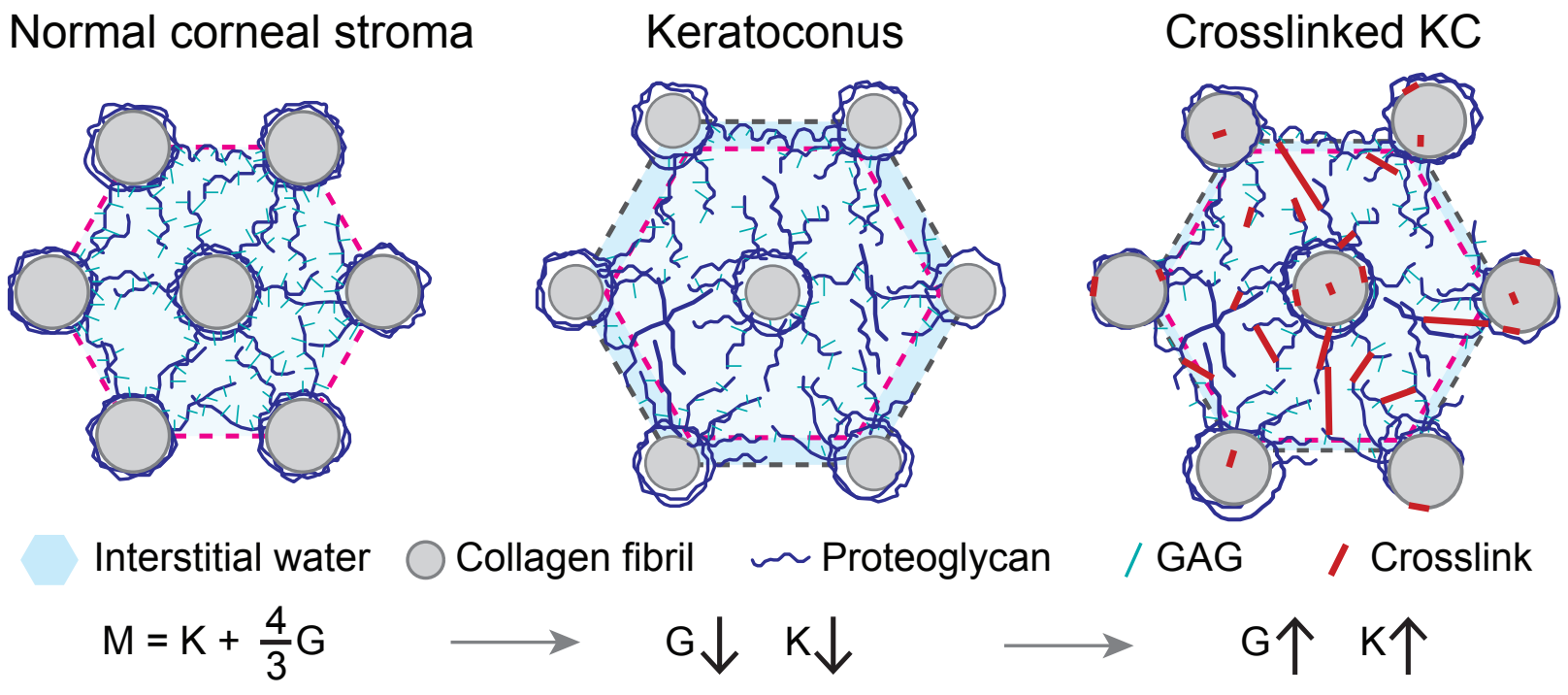

**Fig. 8.** Models of normal, diseased, and CXL-treated corneal tissue and corresponding changes in bulk (K), shear (G), and longitudinal (M) moduli of elasticity. The corneal stroma features hexagonal arrangements of collagen fibrils and interconnecting proteoglycan glycosaminoglycan (GAG) chains (left panel). Transmission electron microscopy showed that collagen fibrils have a uniform diameter of 25-30 nm in normal corneas[61,86], but exhibit reduced diameter in KC corneas[87]. The degradation of collagen fibers and loss of GAG and proteoglycan loosens the microstructural arrangement and leads to a decrease of shear and bulk moduli, and, therefore, longitudinal modulus (middle panel). CXL induces compaction of the structure and promotes regeneration of the constituents[88], consequently increasing shear, bulk, and longitudinal moduli of elasticity (right panel). The changes in longitudinal modulus, which may be spatially heterogeneous, can be probed noninvasively by Brillouin light scattering microscopy.

## Supplementary Information

**Relation of Brillouin frequency shift and longitudinal elastic modulus**

Spontaneous Brillouin scattering is the scattering of light from acoustic phonons (i.e. propagating density waves originating from intrinsic thermal fluctuations) [1,2]. Light scatters from acoustic waves that are phase-matched, resulting in a negative (Stokes) and positive (Anti-Stokes) Doppler frequency shift of the light by the frequency of the mechanical waves. It is the (absolute) magnitude of Brillouin frequency shift that we measure with our optical spectrometer. Although both shear and longitudinal waves are present in the tissue, our Brillouin instrument is configured to measure 180º back-scattering from longitudinal waves only (Fig. 1). In this back-scattering configuration, the Brillouin frequency shift $\Omega$ is related to the acoustic speed V, via $V = \Omega \frac{\lambda}{2n}$, where $\lambda$ is the wavelength of the light, and n is the refractive index of the tissue. The acoustic speed is itself a function of the longitudinal elastic modulus M through the relation $V = \sqrt{M/\rho}$, where $\rho$ is the mass density of the tissue. From this equation, we get $\frac{\Delta\Omega}{\Omega} = \frac{\Delta M}{2M} + \frac{\Delta n}{n} - \frac{\Delta\rho}{2\rho}$. Based on published data [3,4], we estimate $\frac{\rho}{n^2} \approx 0.565$ to 0.5635 g/mL, varying <0.3% within normal corneas. Therefore, the last two terms are nearly canceled out. In general, the acoustic speed has directional dependence and M is a tensor. If we approximate the corneal tissues as a mechanically isotropic material, the relationship between different mechanical moduli is as follows: $G = K\frac{3(1-2\nu)}{(1+\nu)}$; $M = K + \frac{4G}{3}$, where G is shear modulus, K is bulk modulus, and $\nu$ is Poisson's ratio. K is an inverse of volume compressibility. In tissues, M is approximately equal to K. The typical value of longitudinal modulus is several GPa whereas that of shear modulus of tissues is in a 10-100 kPa range [5].

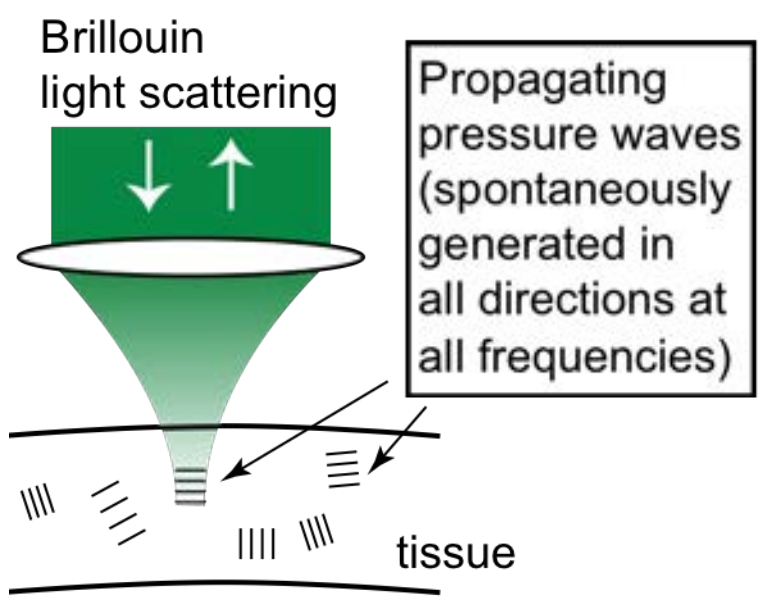

Fig. S1: Brillouin light scattering in the cornea.

Empirically, we have previously observed a relatively high correlation between longitudinal modulus (M) and quasi-static shear modulus (G) in *ex vivo* porcine corneas (Fig. S2) when fit to a log-log linear curve: $\log(M) = c\log(G) + d$, with fitting parameters $c = 0.033$ and $d = 9.31$. For *ex vivo* samples, the effect of swelling may contribute to the apparent correlation between longitudinal and shear moduli. That is, tissues with low shear modulus tend to swell more, and the increased swelling reduces longitudinal modulus. We do not yet know to what extent the observed empirical relation would hold in the physiological condition *in vivo*, in which tissue hydration is regulated. Regardless, the Brillouin frequency shift itself is a convenient metric directly measurable with Brillouin light scattering spectroscopy. In this paper, we report our measurements as Brillouin frequency shifts in GHz.

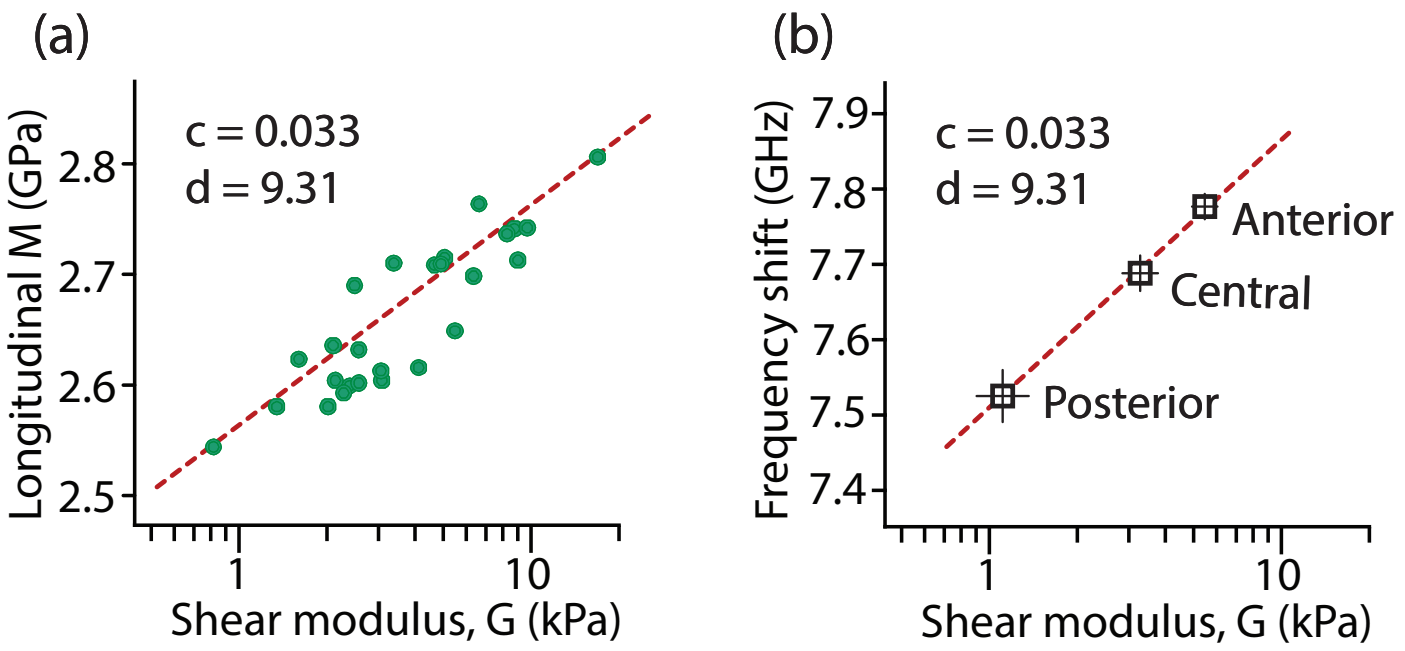

Fig. S2: Empirical quantitative relationship between Brillouin modulus (frequency shift) and shear modulus (G). (a) Comparison data for thin flaps of porcine corneal tissue *ex vivo* (each data point corresponds to one thin flap). Cornea tissue was cut with a biopsy punch to retrieve thin flaps from the anterior, central, and posterior regions. Shear modulus of the thin flaps was measured using a stress-controlled rheometer (ARG2, TA Instruments) at 0.5 Hz with 0.1% strain amplitude. High correlation (R = 0.9) was obtained with $c$ = 0.033 and $d$ = 9.31. (b) Data from (a) grouped by corneal region and plotted versus Brillouin frequency shift measured at a wavelength of 532 nm.

**The magnitude of Brillouin scattering**

The scattering cross section for Brillouin scattering in an isotropic liquid is given by [6]

$$\sigma = \frac{\pi^2 V}{2\lambda^4}\left(\frac{\gamma^2}{K}\right) k_B T(1+ \cos^2\theta)$$

where $\lambda$ is the optical wavelength, $\gamma$ is the adiabatic electrostriction constant, K is the adiabatic bulk modulus (at the Brillouin shift frequency), $k_B$ is Boltzmann constant, $T$ is the temperature, $V$ is the volume of the interaction region, and $\theta$ is the scattering angle. The adiabatic electrostriction constant is a dimensionless measure of the change of the refractive index upon density variation, i.e. $\gamma = \rho(\frac{\partial\varepsilon}{\partial\rho})$, where $\varepsilon$ is dielectric constant and $\rho$ density. To a good approximation, $\gamma$ is equal to $\varepsilon - 1 = n^2 - 1$. The normalized magnitude of scattering is therefore given by

$$S = \frac{\sigma}{A} = \frac{\pi^2 L}{2\lambda^4}\left(\frac{\gamma^2}{K}\right) k_B T(1+ \cos^2\theta)$$

where $A$ and $L$ are the area and length of the interaction region, respectively. We consider a Gaussian probe beam focused to a spot with numerical aperture, NA, in free space. Then, $L \approx \frac{2\lambda}{\pi(NA)^2}$. In an on-axis, backward confocal geometry ($\theta$ close to 180°), only the scattered light within the solid angle given by the numerical aperture is collected. Integrating the angle-dependent term, $(1+ \cos^2\theta)$, over this range yields $\approx 2\pi(NA)^2$. Therefore, the total amount of light collected as a fraction of the laser photons incident on the cornea is:

$$S_{collected} = \frac{2\pi^2}{\lambda^3}\left(\frac{\gamma^2}{K}\right)k_B T$$

For cornea tissues, we use $n$ = 1.376, $K$ = 2.6 × $10^9$ J/m$^3$, $k_B T$ = 4.23 × $10^{-21}$ J (cornea temperature ~33°C), and $\lambda$ = 0.78 × $10^{-6}$ m, and obtain $S_{collected}$ = 5.4 × $10^{-11}$. The scattering efficiency of measurement is independent of NA. It is interesting to find that the collection efficiency is proportional to $\lambda^{-3}$, while the total amount of scattering is proportional to $\lambda^{-4}$ as is well known for Rayleigh and Brillouin scattering.

**Brillouin microscopes**

*Light source*

The light source is a single-frequency, external cavity diode laser (ECDL) with a center wavelength of 780 nm (Vantage TLB-7113, New Focus). Amplified spontaneous emission (ASE) present in the ECDL output is suppressed by approximately 20 dB by passing the laser light through a solid silica, Fabry-Perot etalon (FSR = 15.14 GHz, finesse = ~32, LightMachinery). A small amount of laser light (~4%) is picked off following the etalon and measured with a photodetector so that the ECDL center frequency can be actively locked to a transmission peak of the etalon. The resulting fluctuations in laser intensity after locking are < 5%. Locking the laser to the etalon also reduces wavelength drift of the ECDL output, such that the absolute stability is ± 20 MHz over 5 minutes, limited by temperature-induced/mechanical drift of the etalon itself [7]. An optical isolator (IO-5-780-VLP, Thorlabs) is placed directly following the laser output port to prevent back-reflections from entering the laser.

*Human interface*

Laser light is carried to the human interface via polarization-maintaining single-mode optical fiber (PM780-HP, Thorlabs), where it is collimated and then split into sample and reference arms using a half-wave plate and polarizing beam splitter (CM1-PBS252, Thorlabs). In the sample arm, laser light is focused onto the cornea using a microscope objective (20x, effective NA = 0.1, Mitutoyo). In the reference arm, laser light is focused onto two known reference materials: polystyrene and water at room temperature, through an achromatic doublet lens (f = 30mm, Thorlabs). The temperature of the reference materials is monitored and saved in real-time so that temperature calibration can be applied to the reference spectrum. In both sample and reference arms, a quarter-wave plate is inserted so that back-scattered light is transmitted through the output port of the polarizing beam splitter, where it can then be coupled into a single-mode optical fiber (780HP, Thorlabs) for transmission to the spectrometer. The human interface is mounted to a stripped-down slit lamp headrest with joystick-based spatial manipulator allowing for manual adjustment of the lateral scanning location on the cornea. The objective lens is mounted on a motorized translation stage (T-LSM025A, Zaber Technologies), allowing for automated axial scanning through the corneal depth. At normal incidence, specular reflection significantly diminishes the signal-to-noise ratio of the Brillouin spectrum. Thus, it is necessary to tilt the human interface by about 15° with respect to the head rest, and additionally have the subject fixate their gaze on a target making an angle of about 10° (in the opposite direction) with respect to the head rest, so that the incidence angle is > 10° over a central zone of ~∅6 mm laterally across the cornea.

For eye tracking, a light emitting diode (940 nm, ~2 mW on sample surface, M940L3, Thorlabs) is used. The near-infrared range is chosen because it reduces the sensitivity of intensity-based pupil-detection algorithm to eye color. The eye is imaged using a high-resolution, monochrome CMOS camera (Mako G-419B, Allied Vision). A long-pass filter is used to reject the laser line and background room lights. Frames are captured on the CMOS camera synchronously with the EMCCD (but at a higher frame rate) so that events such as blinking or eye movement can be identified. Because the relative positions of the objective lens and CMOS camera are fixed, the laser location on the CMOS camera is stationary. Hence, the lateral coordinates for each axial Brillouin scan be determined relative to the pupil center by detecting the pupil location in each CMOS camera image.

*Spectrometer*

The spectrometer design is based on two virtually imaged phase array (VIPA) etalons in a cross-axis configuration, similar to that reported previously [8,9]. The two VIPAs used are identical (R1 = 99.9%, R2 = 95%, tilt angle ≈ 1.5°, LightMachinery). A series of lenses (cylindrical lenses C1 and C2, plus spherical lenses A1 and A2) with focal length 200 mm are used to focus light onto the VIPAs, where angular dispersion occurs, and to relay light to two spatial filters (S1 and S2), which block undesired spectral components (Rayleigh peaks). The gradient intensity filter (G) following the first VIPA (V1) enhances the spectral contrast by reshaping the intensity profile along the dispersion direction [10]. The first VIPA (V1) is oriented vertically, while the second VIPA (V2) is oriented horizontally, resulting in a net frequency dispersion along the diagonal axis, which is imaged onto a diagonally-oriented, electron-multiplying charge-coupled detector (EMCCD, iXon 897, Andor Technologies) using a telescope (T). In the portable system, additional folding mirrors are used to make the spectrometer more compact. Typical Brillouin signal counts detected at the EMCCD are on the order of 900 photons/s, requiring exposure times of 100 – 400 ms. The rate of spectrum dispersion on the EMCCD is ~0.125 GHz/pixel.

Table S1. Specifications of the Brillouin Systems

| Parameter | Value |
|---|---|
| Laser wavelength | 780 ± 0.2 nm |
| Laser power on the cornea | 3 – 5 mW |
| Data acquisition time per point | 0.2 – 0.4 s |
| Transverse resolution | ~ 5 µm |
| Axial resolution | ~ 35 µm |
| Spectral sensitivity | ~ ± 8 MHz |
| Measurement repeatability | ~ ± 10 MHz |

**Spectrometer frequency calibration**

To calibrate the spectrometer, we used two materials with known Brillouin shifts, water and Polystyrene (PS). Frequency shifts of the two materials at room temperature were pre-determined with calibrated spectrum dispersion, as $\Omega_{Water}$ = 5.11 GHz and $\Omega_{PS}$ = 9.62 GHz, respectively. Spectral dispersion rate is then determined by:

$$\mathrm{SD} = 2 * \frac{\Omega_{PS} - \Omega_{water}}{(P_{PS-B} - P_{PS-A}) + (P_{W-B} - P_{W-A})}$$

where $P_{PS-A}$ and $P_{PS-B}$, $P_{W-A}$ and $P_{W-B}$ are paired peaks in the reference signal from PS and water, respectively (Fig. S3d). Precise free spectral range during a measurement is estimated as:

$$\mathrm{FSR} = 2 * \Omega_{PS} - |P_{PS-B} - P_{PS-A}| * \mathrm{SD}$$

Sample frequency shift $\Omega_x$ is defined as the distance along the spectrum line from the Rayleigh peak (laser line, blocked by optical shutter and therefore not visible in this figure) to the fitted signal peak, and is given by (Fig. S3b):

$$\Omega_x = 0.5 * [\mathrm{FSR} - (P_2 - P_1)\,]$$

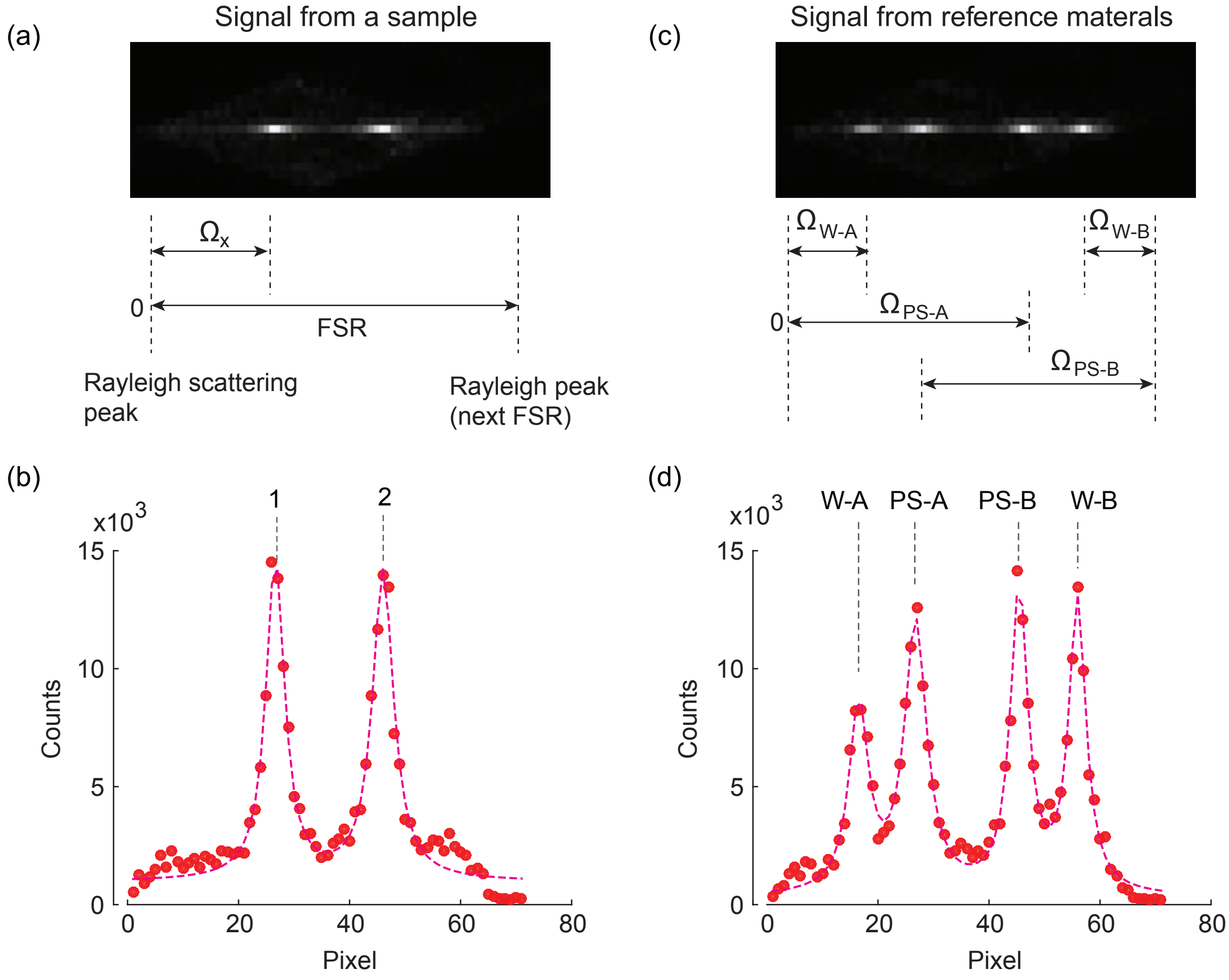

Fig. S3: Spectrometer frequency calibration. (a) EMCCD recording of a typical sample signal. (b) Plot of signal intensity across spectrum line from (a) and resulting fit to a Lorentzian function. Dots: pixel intensity values; dashed line: fitting results. (c) EMCCD recording of a typical signal from reference materials, water and Polystyrene (PS) by focusing light on the water-PS interface. (d) Plot of signal intensity across spectrum line from (c) and the resulting fit to a Lorentzian function. Dots: pixel intensity values; dashed line: fitting results.

**Signal-to-noise ratio (SNR) and measurement sensitivity**

For a CCD detector, shot noise is ideally the primary contributor to signal-to-noise ratio (SNR), with negligible contribution from dark current and readout noise (operation in the so-called, shot-noise-limited regime). However, in detectors using amplifying technologies, such as EMCCDs, the signal multiplying process increases the degree of variance in the signal around the mean value. Technically the detectors work in the shot-multiplicative-noise-limited regime in the best case. For simplicity, we still use the term 'shot-noise-limited regime'. Our spectrometer reading is background free and works in this regime. To verify this, we measured the SNR at different illumination energy levels on a piece of poly(methyl methacrylate) (PMMA) material, using 5 mW input power and integration time increasing from 50 to 600 ms. For each illumination level, 50 measurements were obtained continuously at a single location on the sample. SNR is defined as the ratio of mean signal intensity to standard deviation of signal intensity. A good square root dependence of SNR on the product of illumination power and integration time (total number of photons collected) was found, suggesting shot-noise limited operation (Fig. S4a). In this regime, the measurement accuracy is ultimately limited by the SNR. To find an optimal operating point, we plot measurement errors (standard deviation of measurements) versus 1/SNR. An EMCCD integration time of 0.2 s corresponds to standard deviation of <10 MHz at 5 mW. In Fig. S4c, we show the resulting distribution of estimated Brillouin frequency shifts using these settings (5 mW, 0.2 s) when the measurement is repeated 400 times continuously. The standard deviation of these measurements is ± 7.8 MHz. This corresponds to a relative measurement uncertainty of ~ 0.1 to 0.2%.

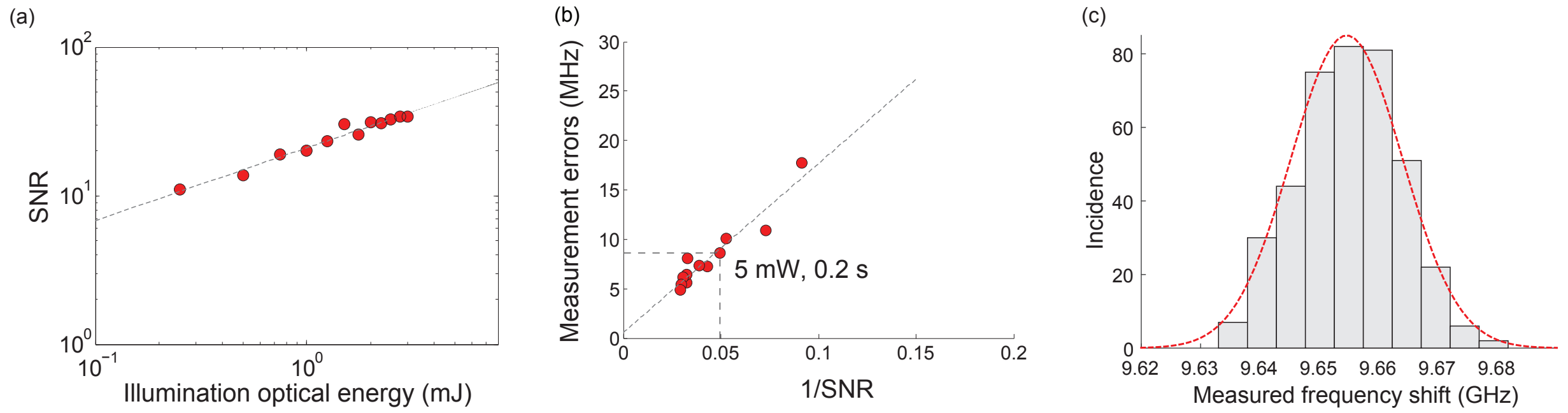

Fig. S4: Signal to noise ratio and measurement sensitivity. (a) Plot of signal-to-noise ratio versus optical energy. Line: linear fit. (b) Plot of measurement error (based on standard deviation of 5 consecutive measurements) versus inverse SNR. Line: linear fit (c) Histogram of estimated Brillouin values over 400 consecutive measurements. Standard deviation is ± 7.8 MHz.

**Spatial resolution**

Light delivery to and light collection from the sample is arranged in a confocal configuration. The foci for light delivery and light collection overlap, hence the spatial resolution is determined by the focal power of the objective lens. Lateral resolution is calculated to be approximately $R_{lateral} \approx 5$ μm. We experimentally determine axial resolution. An edge spread function ESF(z) was obtained via axial scanning though the plastic-water interface of a PMMA cuvette, with a step size of 20 μm. The axial resolution is estimated to be $\mathrm{R_{axial}} = \frac{\mathrm{d}}{\mathrm{dx}}\mathrm{ESF(z)} \approx 35$ μm, given by the FWHM of the fitted Gaussian profile to the ESF (Fig. S5b).

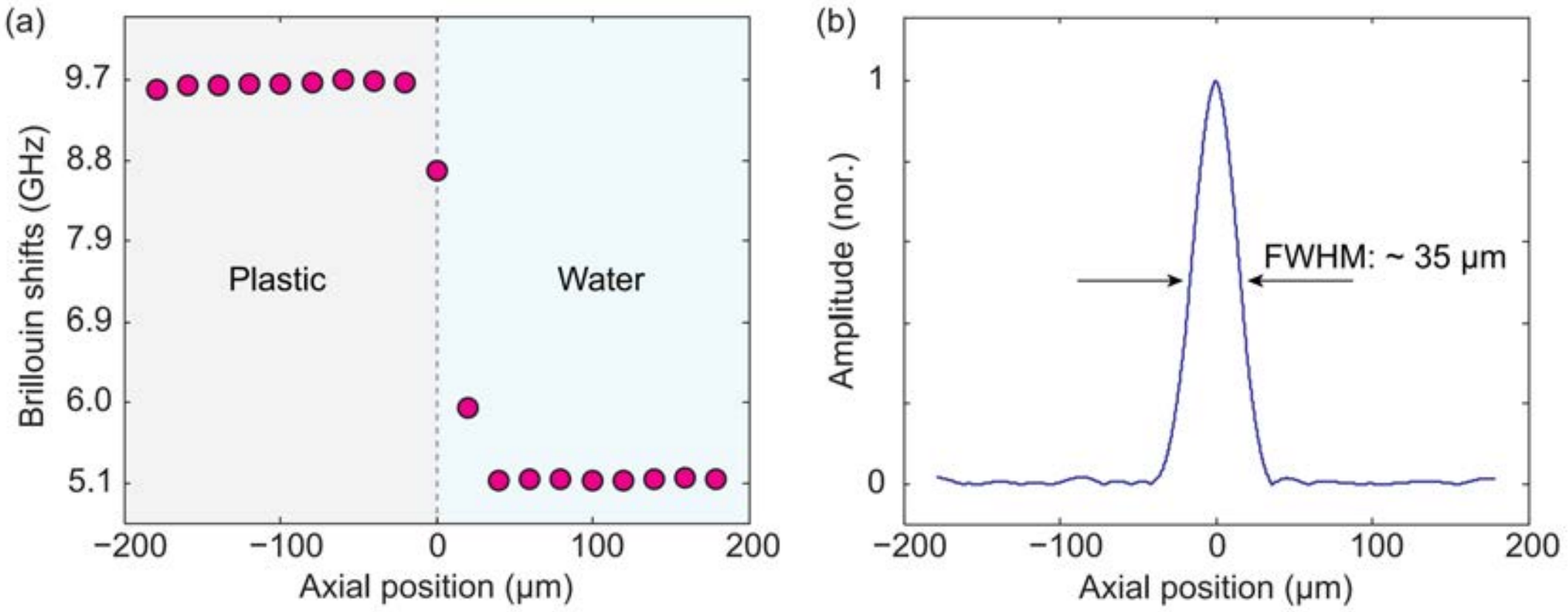

Fig. S5. Axial resolution. (a) Axial scanning profile of a PMMA-water interface. (b) Fitted Gaussian function to the $1^{st}$ order derivative of the interface edge spread function ESF, from which we estimate the axial resolution of the system to be around 35 μm.

## Temperature dependence of Brillouin measurements and calibration

Brillouin frequency shift is temperature-dependent, so temperature fluctuation of the reference materials will cause error in our calibration, and hence our estimation of the Brillouin frequency shift of the sample. We studied the temperature dependence of the reference materials and applied a temperature correction in post-processing to alleviate its effect on measurements. We first measured Brillouin frequency shift in water versus temperature and compared with theoretical prediction (Fig. S6a), calculated using $\Omega = 2n\frac{v_{water}}{\lambda}\sin(\beta)$, where $n$ is the refractive index of water, $v_{water}$ is the speed of sound in water, $\lambda$ is laser wavelength, which in this study was 780 $nm$, and $\beta = 180°$ is the scattering angle. Refractive index and speed of sound in water at different temperatures are taken from literature [11]. Speed of sound is calculated following experimental reports by Bashkatov and Genina [12], which is valid from 0 – 100°C. Distilled water was pre-heated to > 80°C, and data was acquired as the water cooled down to room temperature. 10 consecutive Brillouin measurements were taken with 5 mW input power and ~ 0.7 s integration time at each temperature, which was measured with a high precision thermometer (HH804U, Omega). Temperature drift occurred during in vivo data acquisition. To correct for this, we quantified the temperature-induced measurement variations within a narrower temperature window ranging from 20 – 30 °C (corresponding to the maximum temperature range in a clinical setting), and created a lookup table containing Brillouin frequency shift of the reference materials at different temperatures by fitting 2nd polynomials to the data (Fig. S6b). The table was then used to correct the calibration data taken for each axial scan.

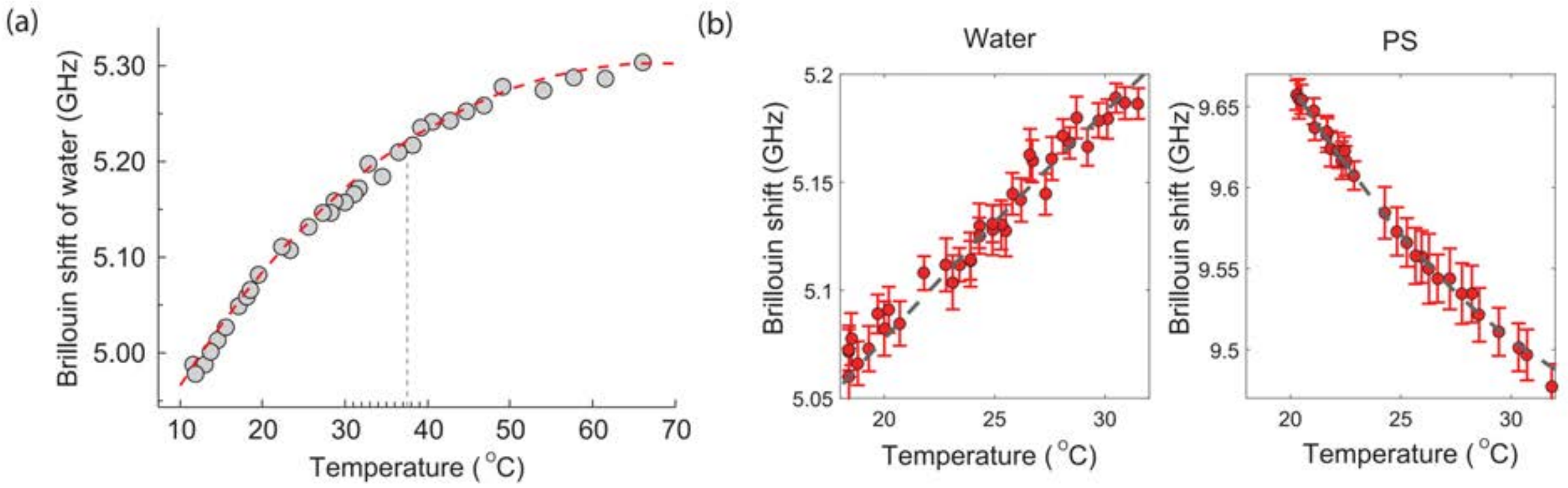

Fig. S6: Temperature dependence of Brillouin measurements and calibration. (a). Brillouin frequency shift of water versus temperature with theoretical prediction. Dots: raw data; red dashed line: prediction. (b) Temperature dependence of Brillouin frequency shift in water and PS over the 20 – 30°C window for temperature correction.

## Brillouin measurement short-term and long-term stability

Both short-term and long-term measurement instability may be caused by environment-induced light source and/or spectrometer drift, mechanical instability of the optical system, etc. To minimize these effects, the imaging system was calibrated daily to maintain its performance. Experiments were also carried out to quantify the measurement variations of the system. In laboratory conditions with typical temperature drift up to 1°C per hour, we characterized short term measurement stability through 12 consecutive measurements of the Brillouin frequency shift of a water sample, taken at intervals of 2.5 minutes. At each of the 12 time-points, the Brillouin frequency value is itself obtained from the average of 5 measurements. Averaged values lie within a ±5 MHz range above and below baseline (Fig. S7a). To examine day-to-day variation, the Brillouin frequency shift of a PMMA sample was measured over four consecutive days at the same time of day in the laboratory environment. 5 measurements were averaged for each daily Brillouin value. Measurements were obtained with 0.2 s integration time at 5 mW input optical power. Day-to-day variation is found to be < 10 MHz (Fig. S7b). On a longer timescale, Brillouin measurements of water over a 140-day period indicate that the long-term stability is ~ ±10 MHz (Fig. S7c).

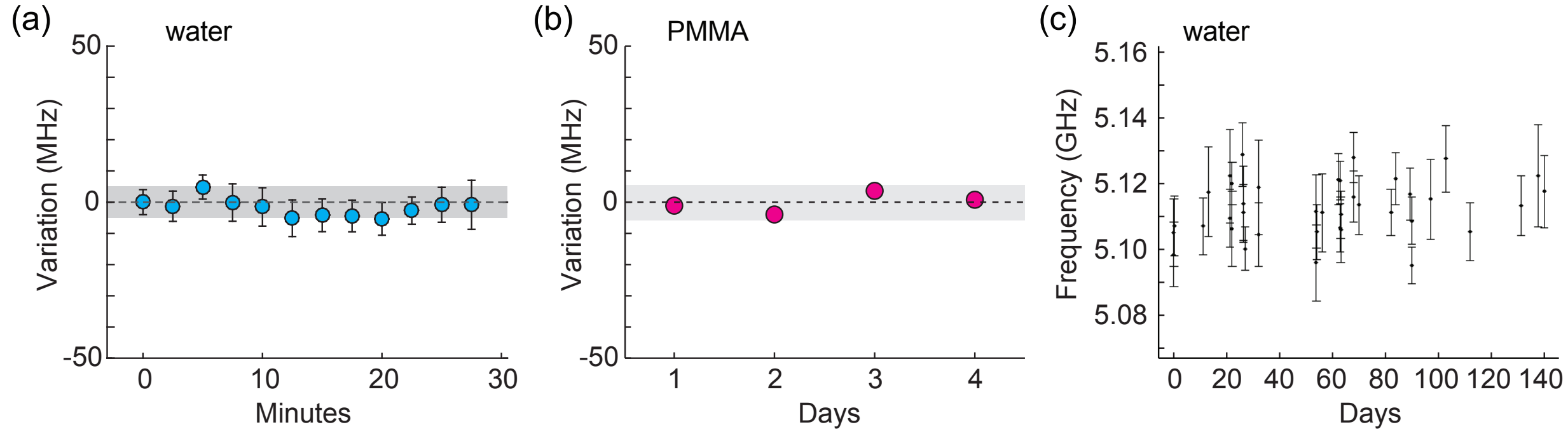

Fig. S7. Measurement of short-term and long-term stability. (a) 12 measurements of water, taken 2.5 minutes apart in the laboratory environment. Dots: mean of 5 measurements; Error bar: standard deviation of 5 measurements; Shadowed zone: ±5 MHz window around the baseline. (b) Brillouin measurements of a PMMA sample over four consecutive days, taken at the same time of day in the laboratory environment. Dots: mean of 5 measurements; Error bar: standard deviation of 5 measurements; Shadowed zone: ±5 MHz window around the base line. (c) Repeated Brillouin measurements of water over a 140-day period, yielding ±10 MHz long-term stability.

## Daily variations in *in vivo* Brillouin measurements

We performed preliminary experiments to assess the daily variation of measurements for individual subjects. For this, we measured Brillouin frequency shift of 3 healthy subjects' corneas over 3 consecutive days at the same time of day (around 5 o'clock pm) in a laboratory with no environmental (temperature, humidity) control. Before Brillouin measurements, central corneal thickness was measured with corneal topography. Standard deviation of CCT measurements of subjects 1, 2, and 3 were: 4.7, 1.0, and 5.5 µm, respectively (Fig. S8a). 5 measurements taken in the central cornea with a radius (R) of < 1 mm from pupil center were averaged to obtain each daily Brillouin frequency-shift value. Standard deviation of the measurements from subjects 1, 2, and 3 were 0.0042, 0.0062, and 0.0169 GHz, respectively (Fig. S8b). Mean standard deviation of Brillouin frequency shift from the three subjects was 0.009 GHz, which is within the sensitivity of our instrument.

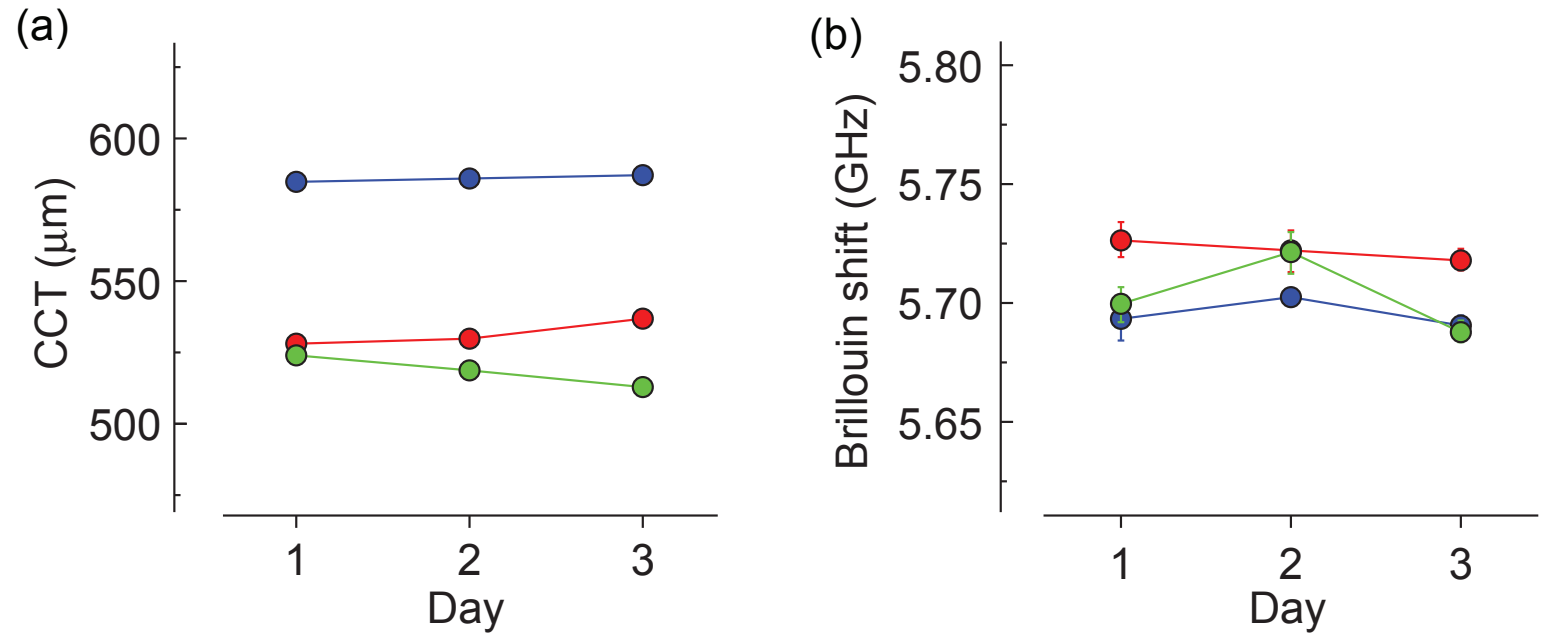

Fig. S8. Daily variation of *in vivo* measurements. (a) Central corneal thickness of three healthy subjects over three consecutive days at 17:00, measured using corneal Scheimpflug topography. (b) Brillouin frequency shift measurements from the same subjects color-coded as in (a). Dots: average of 5 central measurements; error bars: standard deviation of the 5 central measurements.

**Definition of zones for analyzing spatial variation in Brillouin measurements**

To analyze the spatial variation in Brillouin values across the cornea, we define various regions of interest (see Fig. S9 for a graphical illustration of these zones). Central region is defined as the zone within a R = 1 mm radius around the pupil center; Peripheral region in contrast is the area with a distance R > 3 mm from the pupil center. In keratoconic corneas, the 'cone region' is defined as the area with R < 1 mm from the thinnest point (the 'cone center') identified with Scheimpflug-based corneal topographic imaging, and the 'outside-cone region' refers to the region R > 3 mm from the cone center.

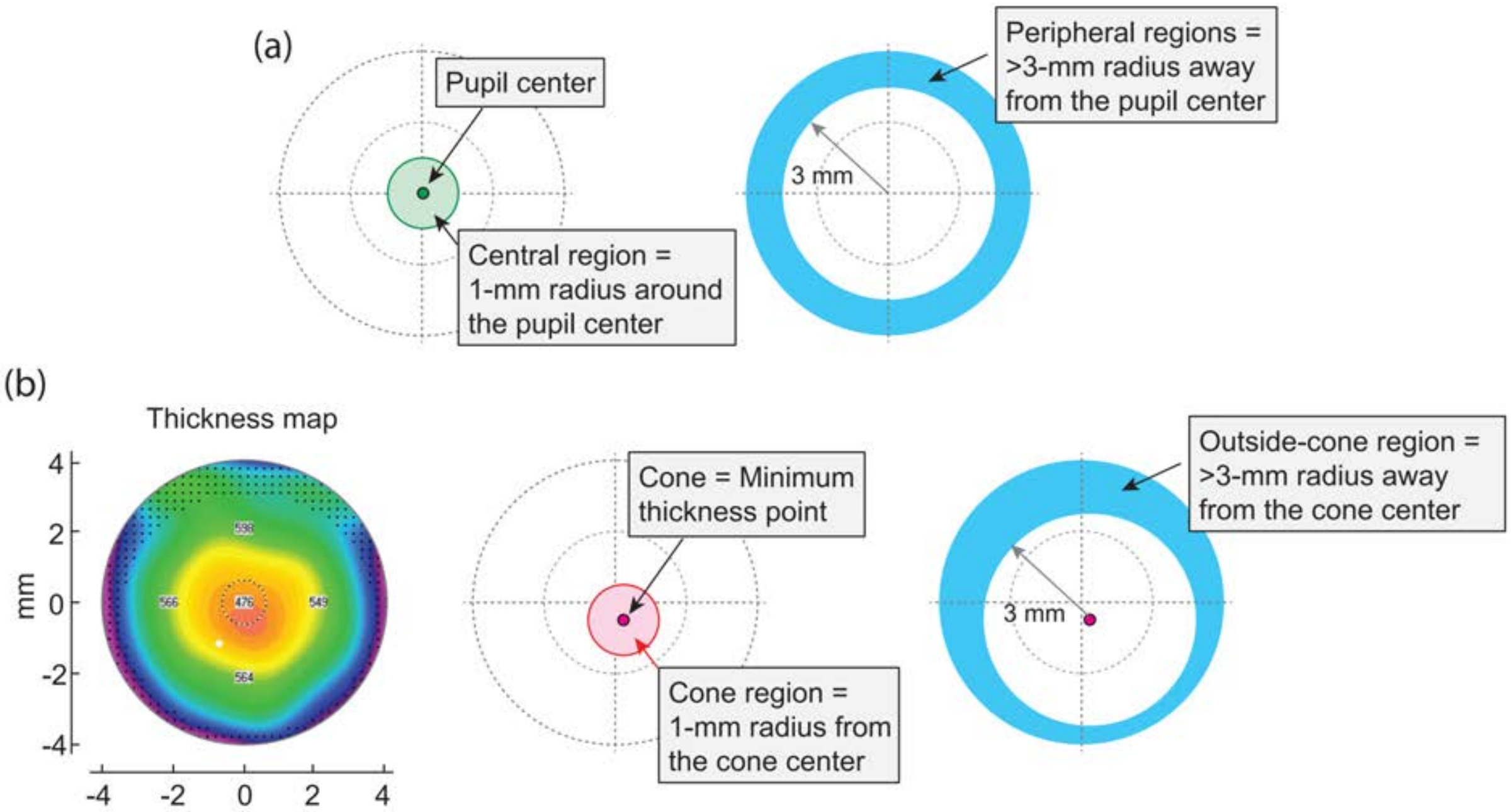

Fig. S9. Definition of zones for analyzing spatial variation. (a) Central region and the peripheral regions, which are defined as areas on the corneal surface within R < 1 mm and R > 3 mm, respectively. (b) The 'cone region' and 'outside-cone region', defined as areas within R < 1 mm R > 3 mm from the 'cone center' that is the thinnest point identified by Scheimpflug corneal topography, respectively.

**Regional differences in Brillouin frequency shift measured in keratoconic corneas**

We defined annular regions extending radially outwards in 0.5 mm increments, centered on the cone (again specified by the thinnest point in the corneal topography). The Brillouin frequency shift in the corneal stroma was averaged over the spatial extent of these annuli and then the difference was taken with respect to the value in the cone region (<1 mm from the thinnest point). Results from the normal cornea group, and KC stage I – IV groups are presented in Fig. S10.

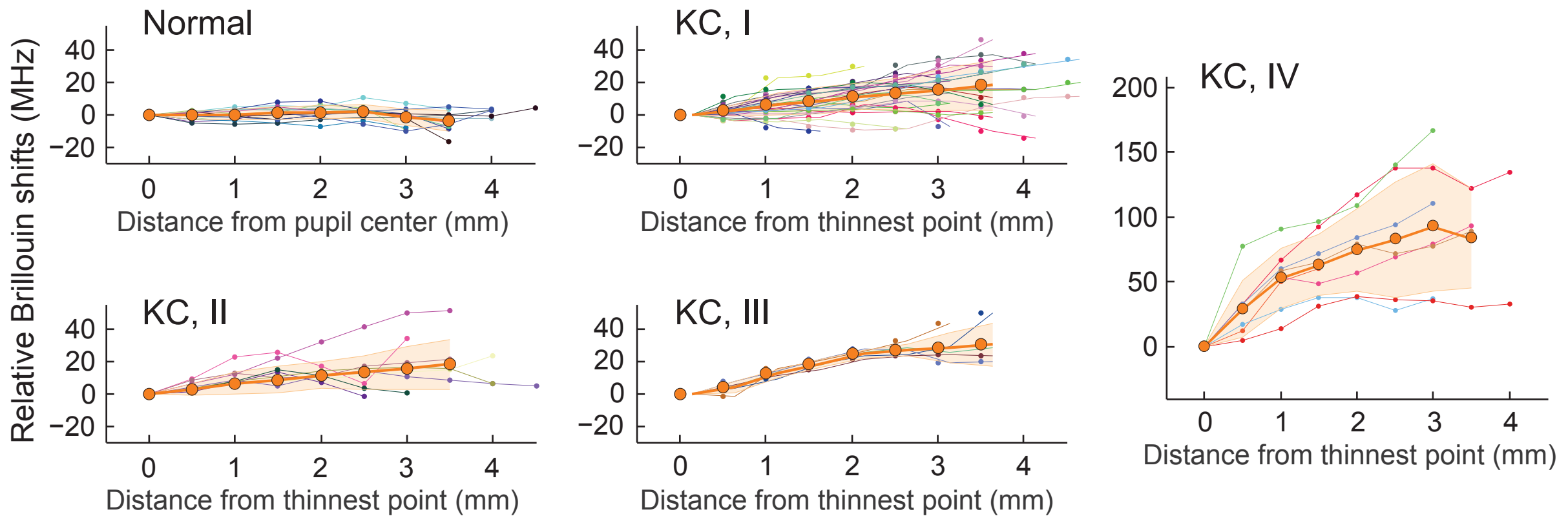

Fig. S10. Regional differences in Brillouin frequency shift measured in corneas. Plots show lateral profiles of average Brillouin frequency shift from cone center to peripheral regions in normal, and KC stage I – IV corneas.

**Safety of Brillouin measurements**

Our Brillouin system uses illumination intensity levels comparable to or less than equivalent ocular instruments that are already FDA-cleared and widely used in clinical practice, such as corneal confocal scanning microscopy and anterior-segment optical coherence tomography. The optical source in our Brillouin instrument emits near-infrared light at a wavelength of 780 nm. Exposure of human eyes to this light during our study was within safety limits established by the American National Standard for Safe Use of Lasers (ANSI Z136.1-2007). The ANSI standard indicates that exposure to 780 nm laser light at the level we use for imaging (5 mW) does not pose a risk to the subject if the continuous exposure does not exceed 5.75 minutes. In our study, the continuous exposure time for each axial scan was < 15 s.

**References:**

[1] L. Brillouin, Ann. Phys. (Paris). **9**, 88 (1922).
[2] J. G. Dil, Reports Prog. Phys. **45**, 285 (1982).
[3] S. Dennis, S. Khan, and K. M. Meek, Biophys. J. **85**, 2205 (2003).
[4] S. Patel, J. L. Alió, and J. J. Pérez-Santonja, Investig. Ophthalmol. Vis. Sci. **45**, 3523 (2004).
[5] B. Fabry, G. N. Maksym, J. P. Butler, M. Glogauer, D. Navajas, and J. J. Fredberg, Phys. Rev. Lett. **87**, 1 (2001).
[6] I. L. Fabelinskii, *Molecular Scattering of Light* (1968).
[7] P. Shao, S. Besner, J. Zhang, G. Scarcelli, and S.-H. Yun, Opt. Express **24**, 22232 (2016).

[8] G. Scarcelli and S. H. Yun, Opt. Express **19**, 10913 (2011).
[9] G. Scarcelli, P. Kim, and S. H. Yun, Opt. Lett. **33**, 2979 (2008).
[10] G. Scarcelli, W. J. Polacheck, H. T. Nia, K. Patel, A. J. Grodzinsky, R. D. Kamm, and S. H. Yun, Nat. Methods **12**, 1132 (2015).
[11] N. Bilaniuk and G. S. K. Wong, J. Acoust. Soc. Am. **93**, 1609 (1993).
[12] A. N. Bashkatov and E. A. Genina, Proc. SPIE **5068**, 393 (2003).